\documentclass[%
aps,
twocolumn,prd,
showpacs,preprintnumbers,
 amsmath,amssymb,
floatfix,
superscriptaddress]{revtex4-1}

\usepackage{graphicx}% Include figure files
\usepackage{dcolumn}% Align table columns on decimal point
\usepackage{bm}% bold math
\usepackage{psfrag}
\usepackage{graphicx}
\usepackage{graphics}
\usepackage{bm}
\usepackage{color}
\usepackage{amssymb}
\usepackage{amsmath}
\usepackage{float}
\usepackage{gensymb}
\usepackage[T1]{fontenc}
\usepackage[latin1]{inputenc}
\usepackage{amsthm}
\usepackage{braket}
\usepackage{lmodern}
\usepackage{fancyhdr} 
\usepackage{rotating}
\usepackage{hyperref}
\newcommand{\gsim}{\raise.3ex\hbox{$>$\kern-.75em\lower1ex\hbox{$\sim$}}}
\newcommand{\lsim}{\raise.3ex\hbox{$<$\kern-.75em\lower1ex\hbox{$\sim$}}}
\newcommand{\be}{\begin{equation}}
\newcommand{\ee}{\end{equation}}
\newcommand{\bi}{\begin{itemize}}
\newcommand{\ei}{\end{itemize}}
\newcommand{\bea}{\begin{eqnarray}}
\newcommand{\eea}{\end{eqnarray}}

\newcommand{\lm}{{\ell,m}}
\newcommand{\NR}{{\rm NR}}

\begin{document}

%\preprint{APS/123-QED}

\title{Comparison of subdominant gravitational wave harmonics between post-Newtonian and numerical relativity calculations and construction of multi-mode hybrids}

\affiliation{ Departament de F\'{\i}sica, Universitat de les Illes Balears and  Institut d'Estudis Espacials de Catalunya,
Cra. Valldemossa km. 7.5, E-07122  Palma de Mallorca, Spain}
\affiliation{Albert-Einstein-Institut, Max-Planck-Institut f\"ur Gravitationsphysik, D-14476 Golm, Germany}
\affiliation{School of Physics and Astronomy, Cardiff University, Queens Building, CF24 3AA Cardiff, United Kingdom}

\author{Juan Calder\'on~Bustillo$^\text{1}$}\email{juan.calderon@uib.es}\noaffiliation
%\email{juan.calderon@uib.es}
\author{Alejandro Boh\'e$^\text{1,2}$}\noaffiliation
\author{Sascha Husa$^\text{1}$}\noaffiliation
\author{Alicia M. Sintes$^\text{1}$}\noaffiliation
\author{Mark Hannam$^\text{3}$}\noaffiliation
\author{Michael P\"urrer$^\text{3}$}\noaffiliation

\begin{abstract}

Gravitational waveforms which describe the inspiral, merger and ringdown of coalescing binaries are usually constructed by synthesising information from perturbative descriptions, in particular post-Newtonian theory and black-hole perturbation theory, with numerical solutions of the full Einstein equations. In this paper we discuss the ``glueing'' of numerical and post-Newtonian waveforms to produce hybrid waveforms which include subdominant spherical harmonics (``higher order modes''), and focus in particular on the process of consistently aligning the waveforms, which requires a comparison of both descriptions and a discussion of their imprecisions. We restrict to the non-precessing case, and illustrate the process using numerical waveforms of up to mass ratio $q=18$ produced with the BAM code, and publicly available waveforms from the SXS catalogue. The results also suggest new ways of analysing finite radius errors in numerical simulations.
\end{abstract}

%check 04.25Dm, dg Nx, 04.80Nn cc, 04.70 Bw, 04.30Db, -w, 07.05 Kf
\preprint{LIGO-P1500004}
\maketitle

\section{Introduction}

Coalescing compact binaries are among the most promising candidates for a first direct observation of the gravitational waves (GW) predicted by Einstein's General Relativity. Within the next few years a new generation of detectors composed by Advanced LIGO \cite{Abbott:2007kv, 2010CQGra..27h4006H, advLIGO}, Advanced VIRGO \cite{Accadia:2011zzc} and KAGRA \cite{Somiya:2011np}, are expected to come online with design sensitivities ten times larger than the previous generation (LIGO and VIRGO). The detection and accurate identification of 
compact binary coalescence (CBC) events relies on the accuracy with which the expected signals are modeled.  
The early inspiral stage of a CBC is well modeled by means of the post-Newtonian (PN) approximation \cite{Blanchet2014}, which consists of a weak-field slow-motion expansion of General Relativity in powers of $GM/rc^2$, or equivalently $v^2/c^2$. However, as the binary tightens, the PN approximation becomes less accurate. In order to model the late inspiral one needs to solve the full Einstein equations, which in turn requires the use of numerical methods. An overview of the capabilities and techniques of numerical relativity (NR) as applied to black hole coalescence is given in \cite{Ajith:2012az,Hinder:2013oqa}. When the final state is a black hole, the perturbed Kerr black hole resulting from the merger will settle down emitting exponentially damped gravitational waves known as quasinormal modes, whose complex frequencies can be computed analytically in linear perturbation theory \cite{Kokkotas:1999bd}, while the amplitudes have to be determined from NR.

In order to synthesize a GW signal that represents all of these stages, typically referred to as an Inspiral-Merger-Ringdown (IMR) waveform, one often needs to align and appropriately glue together the PN and NR descriptions. The construction of such ``hybrid waveforms'' for the dominant quadrupole modes (corresponding to the $\ell=2, \vert m\vert = 2$ spin-weighted spherical harmonic),  which is what CBC searches  have employed to date, has been discussed extensively in the literature, performing the construction both in the time and Fourier domains, see e.g. \cite{OhmeThesis,Santamaria:2010yb}. 
A  publicly available catalogue of such hybrid waveforms is described in \cite{Ajith:2012az}, their use for  calibrating search codes by injecting them into noise in \cite{Aasi:2014tra}. 

The present paper focuses on complete waveforms that also include spherical harmonic modes other than $\ell=2, \vert m\vert = 2$, or ``higher order modes'' (HOM). Our construction will be performed in the time domain (for work in the frequency domain see \cite{Santamaria:2010yb, Damour:2010zb}). For simplicity we will restrict ourselves to non-precessing binaries, i.e.~when the orbital plane is preserved. Hybrid waveforms are essentially constructed in two steps. First we need to align the PN and NR waveforms, taking care of ambiguities of waveforms that describe equivalent physical systems. Then we need to ``blend'' both waveforms together. It is in particular the process of alignment where the presence of more than one spherical harmonic adds subtlety.
The aims of this work are to provide a construction algorithm for waveforms used in injection studies or waveform models including HOM, as well as to evaluate the fidelity of the input and resulting hybrid waveforms, e.g. for constructing waveform catalogues similar to \cite{Ajith:2012az}, which include higher modes.  The type of waveforms constructed here have been already used in \cite{Varma:2014jxa} for the purpose of testing the effect of higher order modes in non-spinning CBC searches. We here present a detailed procedure for such a construction and analyze the main new sources of error that might appear. Our findings regarding the $\ell=2, \vert m\vert = 2$  modes are consistent with previous work, where the influence of several ingredients of the construction (e.g. the choice of the hybridization point [interval], and the shape of the blending functions) has been studied extensively, see e.g.~\cite{Santamaria:2010yb,MacDonald:2011ne,MacDonald:2012mp}.   

Our numerical waveforms have been taken from the publicly available SXS catalogue \cite{SXS} (computed by the SpEC code~\cite{Ossokine:2013zga,Hemberger:2012jz,Szilagyi:2009qz,Boyle:2009vi,Scheel:2008rj,Boyle:2007ft,Scheel:2006gg,Lindblom:2005qh,Pfeiffer:2002wt,Mroue:2012kv,Mroue:2013xna, Buchman:2012dw}),
and from a set of waveforms that have recently been constructed with the BAM code \cite{Bruegmann:2006at,Husa:2007hp}. Our examples will focus on mass ratios 8 and 18, where contributions from higher modes are significantly stronger than for roughly equal masses, but where it is computationally much more expensive to extend NR calculations to low frequencies, where PN is reliable.

The rest of this paper is organized as follows. In Sec.~\ref{sec:notations}, we introduce the basic definitions regarding waveforms and their ambiguities; the basic definitions regarding data analysis which we will need are introduced in Sec.~\ref{subsec:dataanalysis}.
In Section \ref{sec:generaldiscussion}, we discuss the general principles of the construction of  hybrid waveforms. In particular we discuss the three degrees of freedom required to align both waveforms. Section \ref{sec:hybrid22} is devoted to the review of the construction of a single-mode hybrid and the corresponding sources of error are identified and discussed. In Section \ref{sec:hybridhighermodes} we describe the method we followed for coherently constructing our hybrids with higher modes and in Section \ref{sec:errorsources} we study the residual disagreements between PN and NR in the hybridization region after alignment and identify their origin. Finally, in Sec.~\ref{sec:MatchResults} we evaluate the influence of NR extraction radius (and extrapolation) on the hybrid in terms of waveform matches.

\section{Waveform definitions and ambiguities}
\label{sec:notations}

Our goal is to assemble a hybrid waveform from two independently constructed pieces (the early-time part typically computed by post-Newtonian methods and the late-time part computed by solving the Einstein equations numerically), or more generally to compare any two waveforms (such as the results of two numerical relativity calculations). As a first step we then need to understand the different conventions and possible ambiguities that went into the definition of both pieces. We will start defining our waveforms in terms of the Newman-Penrose scalar $\Psi_4$, which is the waveform quantity directly computed in many  numerical relativity codes, and afterwards focus on the strain $h$, which is the quantity directly relevant to the data analysis of current ground-based gravitational wave detectors. $\Psi_4$ is computed by contracting the Weyl tensor $C_{\alpha\mu\beta\nu}$ with the appropriate elements of a suitable null tetrad $\ell^\mu$, $m^\mu$, $\bar{m}^\mu$,  $n^\mu$ (see  \cite{Newman:1961qr} for a detailed description of the formalism). As an example, the BAM code uses the definition 
\be
\label{psi4definNR}
\Psi_4=-C_{\alpha\mu\beta\nu}n^\mu n^\nu\bar{m}^\alpha\bar{m}^\beta,
\ee
where $\ell$ and $n$ are ingoing and outgoing null vectors and $-\ell\cdot n = 1 = m \cdot \bar{m}$. The precise choices made in this code can be found in Sec. III of \cite{Bruegmann:2006at}).

The definition of  $\Psi_4$ carries with it several ambiguities, starting with the overall sign convention for the Riemann and Weyl tensors (including metric signature). As examples, in the BAM code \cite{Bruegmann:2006at}  the conventions from Misner, Thorne and Wheeler \cite{Misner:1974qy} are used, and the opposite sign is used in the SpEC code  (see e.g.~the comment above Eq.~(2.100) of  \cite{ChuThesis}).
In addition, the overall sign in \eqref{psi4definNR} is a convention that may change between different authors.

Furthermore, there is freedom in the choice of the tetrad. While $\ell^\mu$, $n^\mu$ will coincide between different codes in the limit $r\rightarrow \infty$, there is no canonical choice \footnote{See e.g. the discussion below Eq.~(29) \cite{Pfeiffer:2007yz}).} for the complex null vector $m^\mu$ which can be rotated by some angle $\sigma$ ($m^\mu\rightarrow e^{i\sigma}m^\mu)$, leading to a redefinition $\Psi_4\rightarrow e^{-2i\sigma} \Psi_4$.
The different choices in the definition of $\Psi_4$ thus amount to an ambiguity
 $\Psi_4\rightarrow e^{i\psi_0} \Psi_4$, which in physical terms is simply the freedom in defining the two gravitational wave polarizations.

The two real polarizations $h_+$ and $h_\times$ of a gravitational wave can be conveniently represented as a complex strain
\be
\label{eq:complexstraindef}
h(t,\theta,\varphi; \Xi)=h_+(t,\theta,\varphi; \Xi)-i h_\times(t,\theta,\varphi; \Xi) \, ,
\ee
where $t$ is an inertial coordinate at null infinity, $\theta$ is chosen as the angle between the  line-of-sight from the detector to the source and the total angular momentum of the binary (which we choose as our $z$-axis), $\varphi$  is an azimuth angle in the source frame, and the 
intrinsic parameters of the source are collectively denoted as $\Xi$. This quantity can be obtained from $\Psi_4$ by applying a double time integration (see \cite{Reisswig:2010di} for a discussion of the issues arising in this procedure), or directly from projecting the metric perturbation onto some \emph{some} orthonormal polarization triad as is usually done in the PN context. Different choices of triad will again lead to a redefinition of the type $h\rightarrow e^{i\psi_0} h$ (see for instance Eq.~(2.6) of \cite{Arun:2009mc}). 

It is convenient to decompose the strain into spin $-2$ weighted spherical harmonic modes $h_{\lm}$ as
\be
 h(t,\theta,\varphi; \Xi)=\sum_{\ell=2}^{\infty} \sum_{m=-\ell}^{\ell}  Y_{\ell,m}^{-2}(\theta,\varphi) h_{\lm}(t,\Xi),
\ee
where $Y_{\ell,m}^{-2}(\theta,\varphi)$ are the spin -2 weighted spherical harmonic basis functions.

We restrict to the non-precessing case, here
the intrinsic parameters are then the total mass, the mass ratio and the two (dimensionless) 
spin projections, $\Xi=\{M,q,\chi_1,\chi_2\}$ onto the angular momentum of the system.
In this case the geometry is  symmetric with respect to the orbital plane, which is preserved in time. This  equatorial symmetry implies 
\be
\label{eq:alignedspinsymmetryh}
h(t,\theta,\varphi; \Xi)=h^{*}(t,\pi-\theta,\varphi; \Xi)
\ee
(where a ${}^*$ denotes complex conjugation) provided that the polarizations are defined using some appropriate choice for the projection triad/tetrad, which is usually the case in the literature. 
For the individual modes this translates into
\be
\label{eq:alignedspinsymmetryhlm}
h_\lm(t,\Xi)=(-1)^\ell h^{*}_{\ell,-m}(t,\Xi) .
\ee
Therefore, we just need to  focus on the $m\geq0$ modes, except when reconstructing the whole waveforms.
Finally, it is convenient to decompose each mode into a real amplitude and phase as
\be
h_{\lm}(t,\Xi)=A_\lm(t,\Xi)e^{-i\phi_\lm(t,\Xi)}.
\ee
In the following we will omit the dependence on $\Xi$ in order to simplify notation and  write $h(t,\theta,\varphi)$.

We note that during inspiral the phase of the $(\ell,m)$ mode approximately follows the rule $\phi_\lm(t)\approx m \phi_{\rm orb}(t)$, where $\phi_{\rm orb}$ is the orbital phase, however this approximate relation has to break down eventually during the merger, as it is violated during the ringdown, as one can check by comparing the quasi-normal frequencies of the different modes. We will return to this issue in Sec.~\ref{sec:phase_PN_NR} when comparing PN and NR phases of individual modes.

The strain $h^D$ seen by a detector located in the direction $(\theta,\varphi)$ of the source sky also depends on the luminosity distance $d_L$ to the source, and the orientation of the detector with respect to the source, which we parametrize using three angles $(\bar{\theta},\bar{\varphi},\psi)$. Here $(\bar{\theta},\bar{\varphi})$ are the spherical coordinates of the source in the detector sky, and $\psi$ is a polarization angle. This dependence is encoded in the  antenna patterns $F_+$ and $F_\times$ of the detector as
\be
h^D  =\frac{F_+(\bar{\theta},\bar{\varphi},\psi) h_+(t,\theta,\varphi) + F_\times(\bar{\theta},\bar{\varphi},\psi) h_\times(t,\theta,\varphi)}{d_L} .
\ee
where
\bea
F_+&=&\frac{1+\cos^2{\bar{\theta}}}{2}  \cos{2\bar{\varphi}}\cos{2\psi}-\cos{\bar{\theta}}\sin{2\bar{\varphi}}\sin{2\psi},\nonumber \\
F_\times &=& \frac{1+\cos^2{\bar{\theta}}}{2}   \cos{2\bar{\varphi}}\sin{2\psi} +\cos{\bar{\theta}}\sin{2\bar{\varphi}}\cos{2\psi}. \nonumber
\eea
It is well known that this can be rewritten as
\be\label{eq:hthetaphikappa}
h^D = \frac{ F(\bar{\theta},\bar{\varphi},\psi)} {d_L} 
 \left[ \cos \kappa\, h_+(t,\theta,\varphi) + \sin \kappa\,  h_\times(t,\theta,\varphi)\right],
\ee
where $\kappa$ acts as an effective polarization angle and $F/d_L$ is a simple overall amplitude factor.

We can now list the possible ambiguities in the definition of the waveform and its spherical harmonic modes for
two waveforms A and B, computed with different methods and conventions. We use the superscripts A and B to refer to quantities derived from these waveforms. For aligned-spin binaries we assume that computations A and B  preserve the manifest equatorial symmetry of the problem, in particular that the z-axis of the coordinate system we use to define our spherical harmonic mode decomposition is parallel to the angular momentum of the system. The remaining conventions to choose are the origin of the azimuthal angle $\varphi$ of the spherical coordinates,
a polarisation angle $\psi_0$, and the origin of the time coordinate.
Neglecting for the moment issues related to the accuracy of computations A and B, 
 the strains $h^A$ and $h^B$ computed by implementations A and B are related by
\begin{equation}
\label{eqrelationmodesAB}
h^{A}(t,\theta,\varphi)=e^{i\psi_0}h^{B}(t+\tau,\theta,\varphi+\varphi_0),
\end{equation}
where $\psi_0$ and $\varphi_0$ are two angles that encode the different choices in conventions. Of course, the same relation applies to $\Psi_4$. 
As a result, the $h_{lm}$ modes are related by
\begin{equation}
h_\lm^{A}(t)=e^{i(\psi_0 + m \varphi_0)} h_{\lm}^{B}(t+\tau)
\label{generalrelationmodes}
\end{equation}

Usually, conventions are chosen such that  Eq.(\ref{eq:alignedspinsymmetryhlm}) holds for the individual modes. This implies that $\psi_0\in\{0,\pi\}$ and thus
\begin{equation}
h_\lm^{A}(t)=(-1)^{\kappa_0} e^{i m \varphi_0} h_{\lm}^{B}(t+\tau)
\label{generalrelationmodes2}
\end{equation}
with $\kappa_0 \in\{0,1\}$.
In the case where one only considers the dominant $(2,2)$ mode, equations \eqref{generalrelationmodes} and \eqref{generalrelationmodes2} can be rewritten as $h_{22}^A(t)=e^{i\varphi_0'}h_{22}^{B}(t+\tau)$ i.e. the whole angular freedom amounts to a global phase shift. In the multi-mode case, comparing waveforms without ensuring a consistent choice of  $\psi_0$ will lead to incorrect results.

In order to compare or hybridize two waveforms, we thus need to
align them to resolve the above ambiguities, i.e. either keep track of the differences in conventions (which is typically not possible) or infer these from the waveforms themselves. We describe in detail our procedure to do so in Sec.\ref{sec:hybridhighermodes}, which involves defining some notion of distance between the two waveforms and minimizing it over the parameters $(\psi_0,\varphi_0,\tau)$. Note that $\psi_0$ only depends on the difference in the definition of the polarizations between methods A and B and therefore needs to be determined only once whereas $\tau$ and $\varphi_0$ will differ for each pair of waveforms.
Inaccuracies in the waveforms  will in general lead to residual discontinuities between the spherical harmonic modes even after alignment and these are studied in detail in Sec. \ref{sec:errorsources}.

\section{Matched filter analysis}
\label{subsec:dataanalysis}

Gravitational wave signals buried in the noise $n(t)$ of a detector can be extracted using the technique of matched filtering, which is the optimal filtering to extract signals of known shapes buried in % zero-mean, 
stationary Gaussian noise. Much of the complications of gravitational wave data analysis with actual data from detectors arise from non-stationarity and in particular non-Gaussian noise contributions, which we will not consider in this paper.
There is a variety of effects which may cause a calculated waveform to differ from the physically correct one. The relevance of such differences  will depend on the application of the waveforms. It is well known that inaccuracies in the template will in general degrade the performance of the matched filter, and signals may be missed or their parameters (such as masses and spins) incorrectly identified. 
Our analysis of the match of  waveforms will follow the lines of the matched filtering procedure for detecting gravitational waves. We now briefly review some of the data analysis concepts which we will use. 

For time domain real waveforms $h_1(t)$ and $h_2(t)$ one defines the inner product (``Wiener scalar product'') 
\be
\langle h_1 \mid h_2 \rangle = 4\Re \int_{0}^{\infty}\frac{\tilde{h}_1(f)\tilde{h}_2^*(f)}{S_n(f)} df ,
\ee
where $S_n(f)$ is the one-sided power spectral density of the noise $n(t)$, and tildes denote the Fourier transform of the respective time series.
In defining this inner product, we use the fact that the Fourier transform of the real gravitational wave strain $h(t)$ satisfies 
$\tilde{h}(f) = \tilde{h}^*(-f)$, thus allowing us to define the inner product as an integral over positive frequencies only.
In practice, this integral will be performed in a frequency band determined by the detector bandwidth, the lower cutoff frequency $f_0$ being given by the seismic wall of the detector noise. In this paper we consider the predicted 2015-early science run  noise curve for Advanced LIGO with a $30$ Hz  lower frequency cutoff \cite{Aasi:2013wya},  as used in \cite{Aasi:2014tra}, as well as its optimal sensitivity given by the ``zero-detuned high-power'' noise curve \cite{advLIGOcurves}
  with a  $10$ Hz  lower frequency cutoff.

The overlap between two waveforms is defined as their normalized inner product
\be
{\cal O}[h_1,h_2] \equiv \frac{\langle h_1 \mid h_2 \rangle}{\sqrt{\langle h_1 \mid h_1 \rangle \langle h_2 \mid h_2 \rangle}} \, ,
\ee
 and the match as the maximized overlap over a chosen set of extrinsic parameters $\{\lambda_i\}$ of the waveform,
 \be
 {\cal M}[h_1,h_2] \equiv  \max_{\{\lambda_i\}} {\cal O}[h_1,h_2].
 \ee
The extrinsic parameters $\{\lambda_i\}$ may be, for instance, the time of arrival and the coalescence phase of the binary $\{t_0,\varphi_0\}$. As an example, the maximization over $t_0$ is to be understood as the maximum overlap between $h_1$ and $h_2$ obtained by shifting in time one of the templates, i.e., $h(t) \rightarrow h(t+t_0)$. 
In the case of waveforms containing only a single pair of  $(\ell,|m|)$ modes, e.g.~the dominant $(2,2)$ and $(2,-2)$ modes, the maximization over all the extrinsic parameters can be performed analytically as
\bea
{\cal M}[h_1,h_2] &=& \max_{t_0,\theta, \varphi, \psi} {\cal O}[h_1,h_2] = \nonumber \\
&=&\max_{t_0}\bigg{|}\frac{\int_{f_0}^{\infty}\frac{\tilde{h}_1(f)\tilde{h}_2^{*}(f)}{S_n(f)}e^{-i2\pi ft_0}df}{
\sqrt{\langle h_1 \mid h_1 \rangle \langle h_2 \mid h_2 \rangle}}\bigg{|}
\eea
whereas in general the maximization over $\theta$, $\varphi$, and $\psi$ needs to be done numerically.
We will however only consider maximisation over  ${t_0, \varphi_0}$ in this paper.

%%%%%%%%%%%%%%%%%%%%

\section{NR and PN input waveforms}
 \label{sec:generaldiscussion}
\label{sec:sourcesoferror}

\subsubsection{Post Newtonian expansions}

PN expansions compute an approximate solution of Einstein's equations up to a certain order in the expansion parameter. Since only a finite (small) number of expansion terms are known, it is not possible to perform a strict convergence test to estimate the truncation error. In addition,  the approximation breaks down at merger or shortly before. 
Even at a given PN order for the energy and the flux, different treatments in the derivation of the orbital phase from the balance equation give rise to a variety of ``PN approximants'', such as the Taylor approximants  \cite{Damour:2000zb,Buonanno:2009zt}, which are commonly used in gravitational wave data analysis.

The main consequence of the PN truncation error is a phase evolution which progressively deviates from the correct one as the binary evolves. This secular trend translates into the key source of error for the estimation of the time-shift $\tau$ between PN and NR, as shown in Fig.\ref{secular22},  where the secular trend is also shown in comparison with oscillations originating in residual eccentricity of the NR data.
Since secular phasing errors in PN grow with frequency, it is desirable to hybridize at low frequencies, or equivalently with very long NR waveforms, to minimize such errors. Longer NR waveforms are however significantly more expensive computationally. 

In our study we use the Taylor T1 and T4 approximants including 3.5 PN non-spinning \cite{Blanchet2014} and spin-orbit \cite{Bohe2013} and 2PN spin-spin \cite{Mikoczi2005} phase corrections, which we will just denote as T1 and T4 for brevity. We use 3PN non-spinning amplitude corrections for the higher modes \cite{Blanchet2008} and 3.5PN for the 22 mode \cite{Faye2012}. The spin corrections to the amplitudes are known up to 2PN \cite{Buonanno2013}.

\subsubsection{Numerical Relativity}

In NR, the expansion parameter of the PN approximation is essentially replaced by an expansion in the resolution of the numerical grid, and at least in principle it is possible to provide error estimates through a convergence test.  Unless the GW signal is calculated at null infinity, e.g. employing the Cauchy characteristic method \cite{Pollney:2010hs,Reisswig:2010di}, convergence needs to be checked not only with respect to grid resolution but also extraction radius. It is then possible to either  extrapolate to infinity from a series of finite radii (e.g. for the SpEC simulations \cite{Boyle:2009vi,Taylor:2013zia,Szilagyi:2009qz,Scheel:2008rj,Sperhake:2010tu}), or to directly use results from a single finite radius. Furthermore, systematic errors arise due to imperfections of the initial data and initial orbital parameters, in particular  residual eccentricity,  which generates oscillations in amplitude and phase (see e.g.~the discussion in \cite{MacDonald:2011ne}). Unphysical radiation content of the initial data manifests itself in a small GW burst at early times, which is usually referred to as ``junk radiation''.

\begin{figure}[htbp]
\centering
\includegraphics[width=\columnwidth]{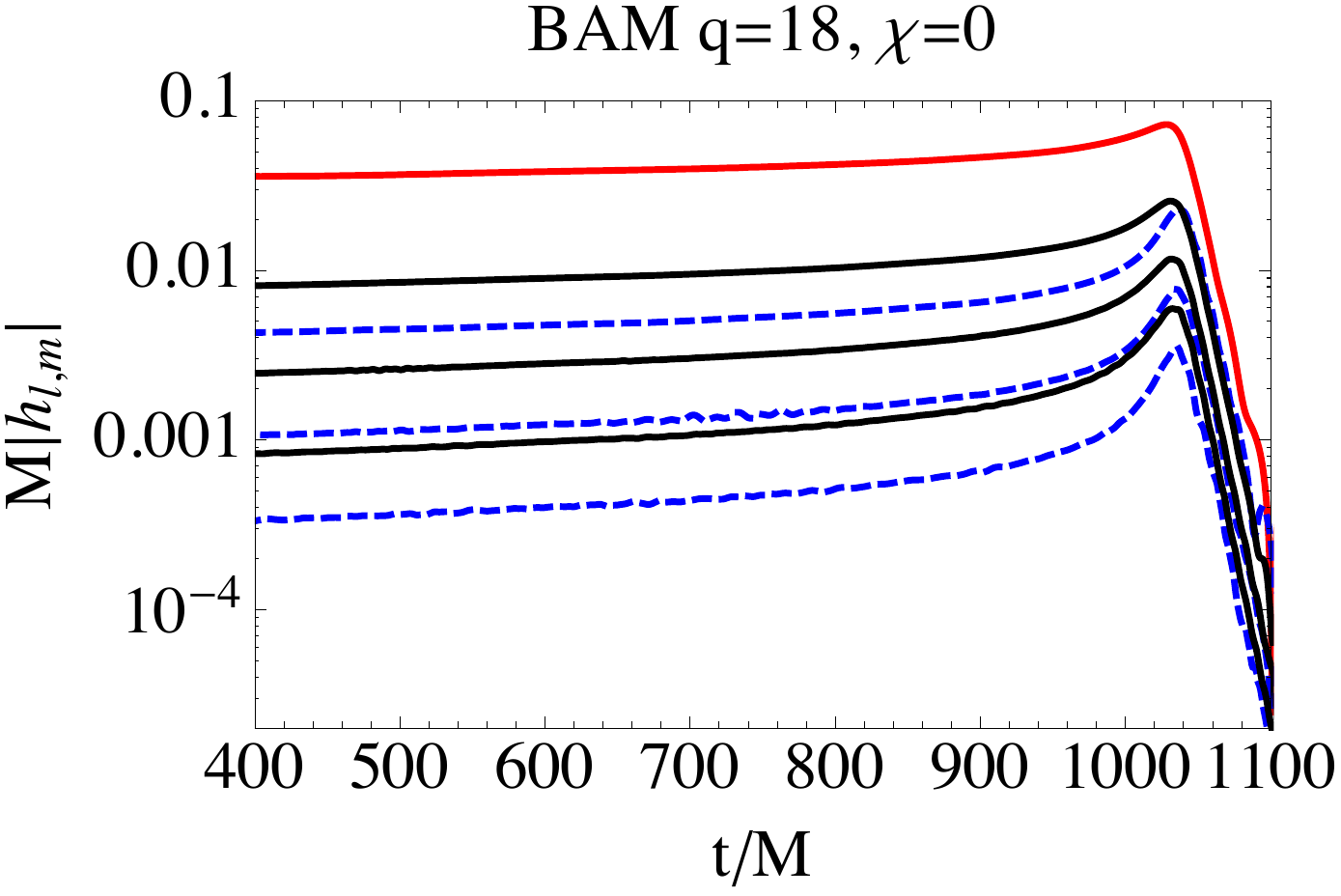}
\caption{Logarithmic plot of the mode amplitudes for the non-spinning $q=18$ configuration. The modes with $(\ell,m=\ell)$ (in black, with $\ell=3,4,5$ from top to bottom) and $(\ell,m=\ell-1)$ (in dashed blue, with $\ell=2,3,4$ from top to bottom) all have a peak amplitude smaller than that of the (2,2) mode (on top) by a factor between $\sim 3$ and 20. We show the clean part of the waveform after initial transients due to junk radiation.}
\label{fig:q18s0ModeAmplitudes}
\end{figure}

For mass ratio $q=18$, new NR simulations have been performed with the BAM code~\cite{Bruegmann:2006at,Husa:2007hp}, and are summarized in Sec.~III.~A of \cite{Varma:2014jxa}.
GW strain is computed from $\Psi_4$ using the fixed-frequency-integration algorithm described in~\cite{Reisswig:2010di}. For a recent comparative discussion of current numerical relativity codes for black hole binaries, including SpEC and BAM see \cite{Ajith:2012az,Hinder:2013oqa}.  Fig.~\ref{fig:q18s0ModeAmplitudes} shows the amplitude of the most important modes for the highest resolution BAM $q=18$ waveform.

\section{Single mode hybrids}
\label{sec:hybrid22}

\subsection{Idealized case}
In order to illustrate some key points of the hybrid construction, we first consider a single mode and assume that we have at our disposal two \emph{infinitely accurate} general-relativity computations of some spherical harmonic mode of the strain, $h^{A}(t)$ and $h^{B}(t)$, that overlap over some portion of the evolution of the binary, i.e. that there is an interval where they satisfy \eqref{generalrelationmodes}. 
Let us define for convenience the amplitude $A(t)$ and the phase $\phi(t)$ of waveform X (with X= A or B) as
$h^{X}(t)=A^X(t) e^{i\phi^X(t)}$ which we assume to be defined over some interval $[0,t_f^X]$, as well as the frequency $\omega^X(t)={d \phi^{X}}/{dt}$ which is a monotonic function of $t$ in the case of binaries on circular orbits (we will discuss below the problems introduced by the oscillations due to some residual eccentricity in the NR waveform; however, provided that the eccentricity is small enough, this remains true). We can therefore  define the inverse function $t^{X}(\omega)$, which satisfies $t^{X}(\omega^X(t))=t.$

Our idealized infinitely accurate waveforms will satisfy
\begin{equation}
\label{ideal22rel}
h^{B}(t)=e^{i\varphi_0}h^{A}(t+\tau)
\end{equation}
for some $\tau$ and $\varphi_0$, and $t$ in the interval $[t^{A}(\omega^{B}(0)),t_f^{A}]$, which implies
%\begin{equation}
$
\omega^{B}(t)=\omega^{A}(t+\tau).
$
%\end{equation}
Determining $\tau$ and $\varphi_0$ is then trivial: one chooses any frequency $\omega_0$ inside the range $[\omega^{B}(0),\omega^{A}(t_f^{A})]$ and obtains
\begin{equation}
\tau = t^{A}(\omega_0)-t^{B}(\omega_0),\quad
e^{i \varphi_0} = \frac{h^{B}(t^{B}(\omega_0))}{h^{A}(t^{A}(\omega_0))}.
\end{equation}
In this idealized case the time alignment and angle $\varphi_0$ do not actually depend on the frequency $\omega_0$ and no blending is required, both functions perfectly overlapping before and after the matching point.

\subsection{Realistic case}

In practice both computations are affected by errors, and \eqref{ideal22rel} is never exactly satisfied over any interval. One rather has to find the best parameters $\tau$ and $\varphi_0$ so that \eqref{ideal22rel} is the closest to being satisfied in \emph{some} sense and over \emph{some} matching window. We thus now have to make some particular choices in our hybrid construction. We parameterize our window by the initial time $t_0$ or initial frequency $\omega_0$ (defined as $\omega^{PN}(t_0)=\omega_0$) and the length of the window in the time domain $\Delta t$, i.e. our window is $[t_0,t_0+\Delta t]$.
In order to avoid the influence of amplitude errors, we only take into account phase information when aligning the waveforms in time. We adopt the quantity 
\begin{equation}
\Delta(\tau; t_0,\Delta t)=\int_{t_0}^{t_0+\Delta t} \left( \omega^{PN}(t) - \omega^{NR}(t-\tau)\right)^2 dt,
\label{eq:omegaintegral}
\end{equation}
which has the advantage of not depending on $\varphi_0$. Other authors have replaced $\omega$ by $\phi$ or $h$ \cite{MacDonald:2011ne} in the integrand (which then depends on $\varphi_0$) and tested that their hybrid was not affected much by this choice.

The appropriate time shift $\tau$ for a given choice of window $[t_0,t_0+\Delta t]$ is then obtained by minimizing $\Delta(\tau; t_0,\Delta t)$. Once this is done, the optimal phase shift $\varphi_0$ has to be determined.  Simple choices are to align the phases at a fixed time, e.g.~the beginning of the window, $\varphi_0=\phi^{NR}(t_0-\tau)-\phi^{PN}(t_0)$, or to pick the phase shift that minimizes $\int_{t_0}^{t_0+\Delta t} \left( \phi^{NR}(t-\tau)-\phi^{PN}(t) + \varphi_0\right)^2 dt$. We have checked that the resulting hybrid depends very weakly on this particular choice. This is due to the fact that the phase has one additional integration with respect to the frequency, so it contains less oscillations. 

Once $\tau$ and $\varphi_0$ have been determined, both waveforms are combined into a piecewise definition
\begin{widetext}
\begin{align}
 h(t)=\left\{
 \begin{array}{cl}
 e^{i\varphi_0}h^{PN}(t+\tau) & \text{ if } t<t_0-\tau\\
w^-(t)e^{i\varphi_0} h^{PN}(t+\tau) + w^+(t) h^{NR}(t) & \text{ if } t_0-\tau<t<t_0-\tau+\Delta t\\
 h^{NR}(t) & \text{ if } t_0-\tau+\Delta t<t
 \end{array}
 \right.
\label{eq:hybrid22}
 \end{align}
 \end{widetext}
where, with the notation $w_{[t_1,t_2]}^\pm(t)$ for blending functions that monotonically go from 0 to 1 (or from 1 to 0) in the interval $[t_1,t_2]$, we have defined $w^\pm(t)=w_{[t_0-\tau,t_0-\tau+\Delta t]}^\pm(t)$, i.e. we perform the blending over the same interval we used to determine $\tau$ and $\varphi_0$. Here again, different authors have made different choices for the exact shape of these functions. For instance, \cite{MacDonald:2011ne} considers cosine functions, while we here use linear ones. 

Note that multiplying the PN part by $e^{i\varphi_0}$ is a redefinition of the conventions for the PN waveform (change of the orbital phase): now the early part of the hybrid does not exactly reduce to the original PN waveform. In the single mode case this is trivial 
(and multiplying the NR part by $e^{-i\varphi_0}$ would have been equivalent). As single mode matches are always optimized over coalescence phase, this redefinition of conventions will never have any practical consequence. As we will see, in the multimode case, things get more involved.

\begin{figure}[htbp]
\centering
\includegraphics[width=\columnwidth]{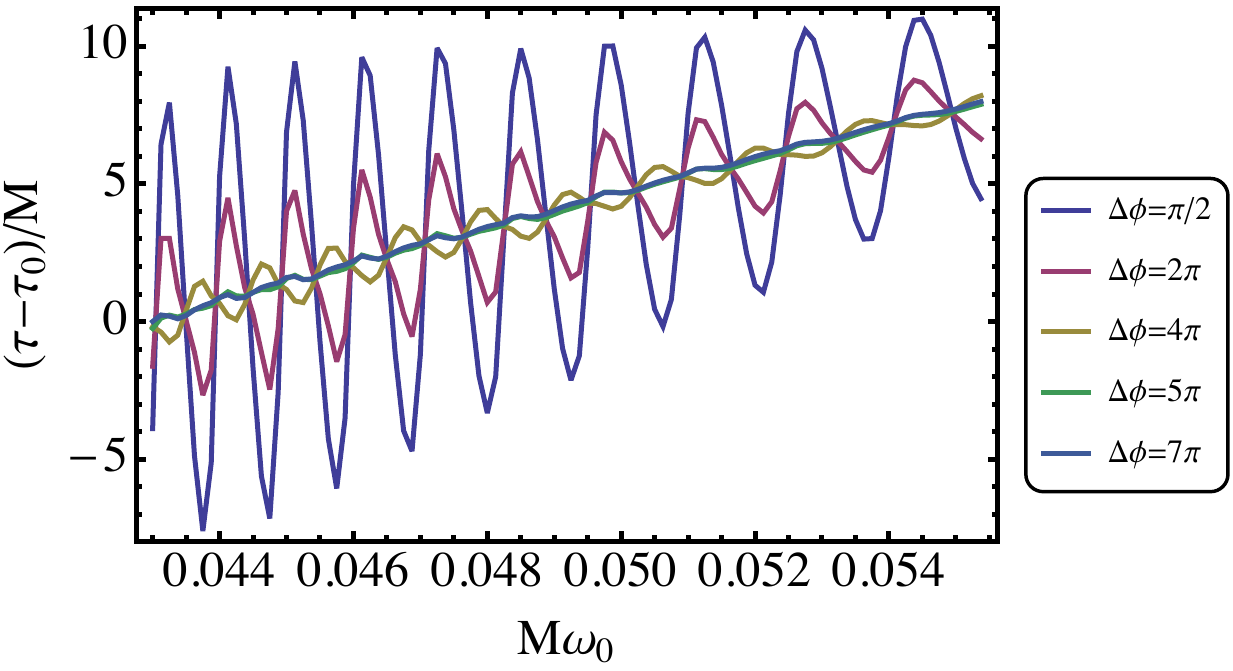}
\includegraphics[width=\columnwidth]{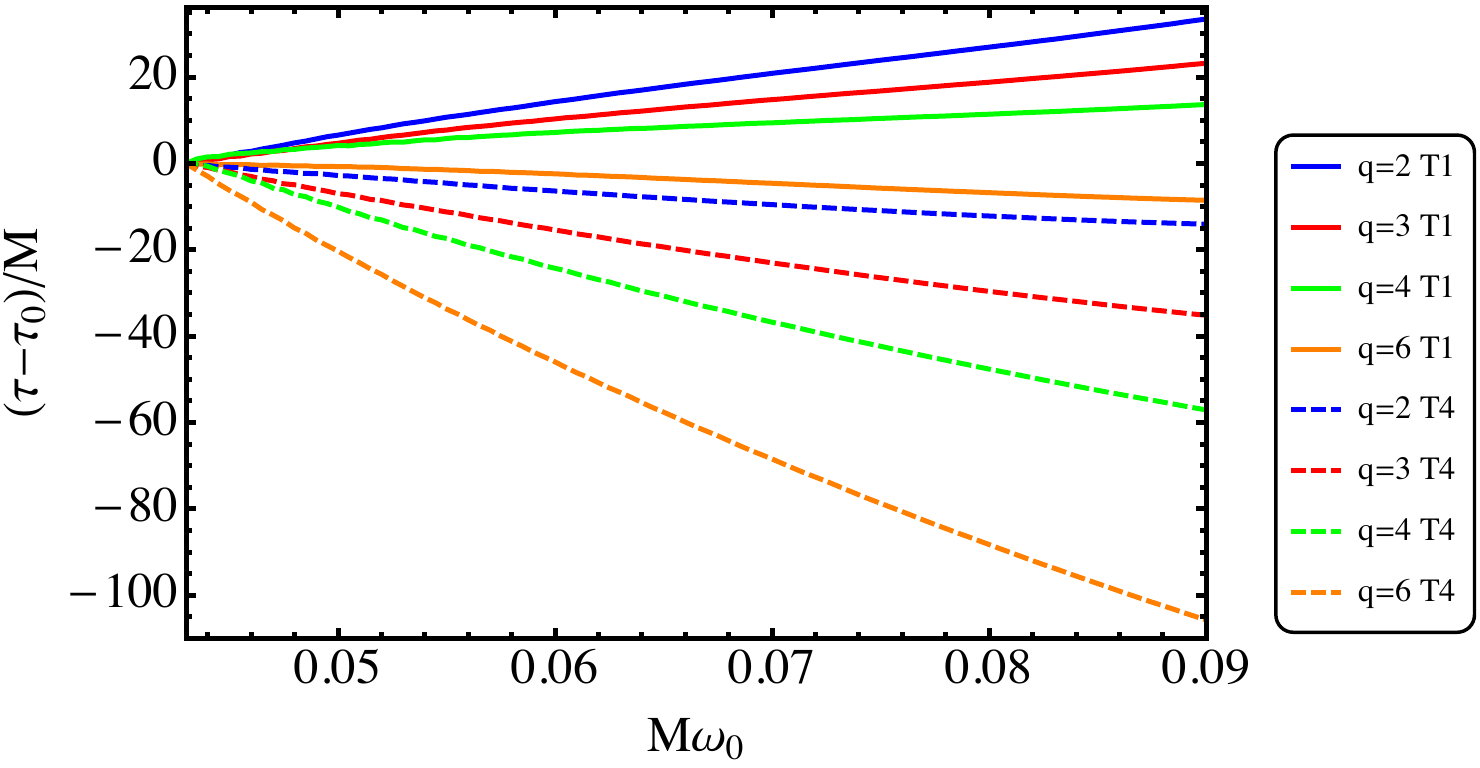}
\caption{Top: relative time shift $(\tau-\tau_0)/M$ as a function of $M\omega_0$ for $q=3$ non-spinning SXS NR data hybridized to  T1 for various lengths $\Delta \phi$ of the matching window. The reference time shift $\tau_0$ is the one obtained for $\Delta \phi=7\pi$ and $M\omega_0=0.043$. Note how for larger $\Delta \phi$ the oscillations in the estimation of $\tau/M$ are smaller. Bottom: same for several SXS data sets hybridized to  T1 and T4 using $\Delta \phi=7\pi$ (smaller values of $q$ on top) and $\tau_0$ as above.}
\label{secular22}
\end{figure}

In order to quantify the error in the time alignment of the waveforms, it is instructive to look at the time shift $\tau$ as a function of matching frequency $\omega_0$ and window size $\Delta t$ (note that the absolute value of $\tau$ for some $\omega_0$ is meaningless, what matters is its variation).
Fig.\ref{secular22} illustrates how our best choice for $\tau$ varies with our choices of window length $\Delta t$, and how the secular trend depends on the choice of PN approximant. 
In the present case, oscillations in the NR waveform are caused in particular by residual eccentricity, which manifests itself at frequencies of the order of the orbital frequency (close to half the frequency of the $(2,2)$ mode).
Indeed,  for $\Delta t$ significantly larger than the orbital period, we see that most of the oscillations in the NR data average out, and for a $\Delta t$ corresponding to at least  $\Delta \phi=\phi^{PN}(t_0+\Delta t)-\phi^{PN}(t_0) \sim 5\pi$ oscillations are smaller than the secular trend due to the phase evolution not being accurate. 
Unless specified otherwise, we therefore choose $\Delta t$ such that $\Delta \phi= 7\pi$, i.e. $3.5$ GW cycles. Another possibility, proposed in \cite{Taracchini:2012ig} is to force the window extremities to lie at some maximum of the modulation to ensure the cancellation of the effects due to the modulation over the window. 

In order to minimize alignment errors due to the secular dephasing between PN and NR,  the interval over which one aligns the waveforms should be chosen as early as possible since the accuracy of the PN perturbative treatment degrades as the frequency increases, but not as early as to be affected by junk radiation or other early-time transient errors. In addition, a comparison of PN approximants as in Fig.\ref{secular22}  can be used to choose a PN waveform with smaller error in the matching region.

\begin{figure}[htbp]
\centering
\includegraphics[width=\columnwidth]{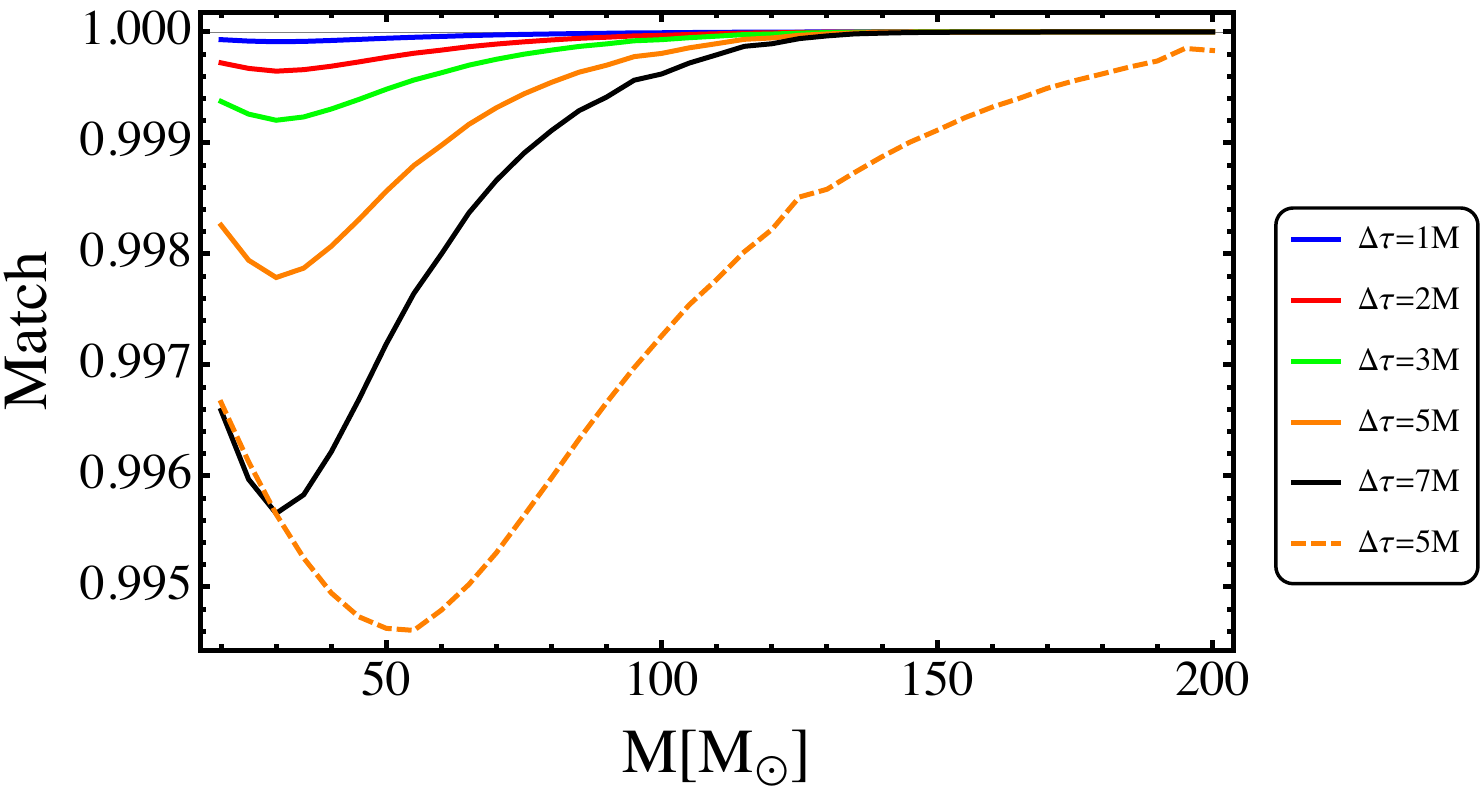}
\caption{Effect of $\Delta \tau$ on the final hybrid $(2,2)$ mode. We show  the match ${\cal M}[h_{\tau_0},h_{\Delta\tau}]$ optimized over time and phase between hybrids built from SXS $q=3$ non-spinning data hybridized to  T1 at $M\omega_0=0.043$ (solid) and $0.073$ (dashed).  An artificial time-shift $\tau_0+\Delta\tau$ is applied when constructing $h_{\Delta\tau}$, while $h_{\tau_0}$ has been built using the optimal time-shift $\tau_0$  between PN and NR. See main text for details.}
\label{arbitrarytimeshifts}
\end{figure}

Now we address the question of how much this impacts the waveform in terms of quantities useful for data-analysis. Fig.~\ref{arbitrarytimeshifts} shows the match between $q=3$ non-spinning hybrid waveforms constructed using artificial time shifts $\tau=\tau_0 + \Delta \tau$ with a reference waveform for which $\tau=\tau_0$. We use the Zero-Detuned High Power noise curve of Advanced LIGO \cite{advLIGOcurves} with a lower frequency cutoff of $10$Hz in order to facilitate comparison with \cite{MacDonald:2011ne}. 
Regardless of the intrinsic parameters of the system and the hybridization frequency, the mismatch increases with $\Delta \tau$. When the hybridization region is in band, the match decays by a few $10^{-3}$ for a value of $\Delta \tau$ of a few $M$, which is consistent with the results 
obtained by \cite{MacDonald:2011ne}. Note that $\Delta \tau$ has a  larger effect on the match
for larger frequency $\omega_0$, as expected from the fact that $\Delta \tau$ is then a larger fraction of the period.

\section{Multi-mode hybrids}
\label{sec:hybridhighermodes}

In this section, we describe our procedure to construct hybrid waveforms with higher modes and define quantities that will be used in the error analysis of Sec.\ref{sec:errorsources}.
Higher modes become increasingly relevant for binaries with large mass ratio, for this reason we illustrate our procedure using a waveform produced recently with the BAM code for a non-spinning binary with $q=18$, which we have presented in Sec.~\ref{sec:sourcesoferror}.

\subsection{Step 1: Determination of $\tau, \psi_0, \varphi_0$}
As discussed in Section \ref{sec:notations}, in the presence of higher modes one needs three parameters $(\tau, \psi_0, \varphi_0)$ to describe the possible differences in conventions between the PN and the NR waveforms. The best choice of such parameters will depend on the matching time or frequency, but it appears fruitful to not make different choices for different directions in the source sky (it seems conceivable but not practical to do this
%in order to compensate for numerical relativity finite radius effects
). 
Several strategies to infer these parameters from the waveforms can be explored. One important ingredient is how to weight the contribution of different modes in determining $(\tau, \psi_0, \varphi_0)$. One natural choice, pursued in \cite{Varma:2014jxa}, is  to define a single set of $(\tau, \psi_0, \varphi_0)$ by minimizing the quantity 
\be
\int {\rm d} t \sum_\lm |h_\lm ^{NR}(t-\tau)e^{i(\psi_0+m\varphi_0)}-h_\lm ^{PN}(t)|,
\ee
where the integral is performed over some window corresponding to the hybridization region, and the contribution of each mode is naturally weighted with its amplitude. Note that \cite{Varma:2014jxa} does not restrict $\psi_0$ to belong to the set $\{0,\pi\}$, and the resulting modes do not in general follow the usual relation \eqref{eq:alignedspinsymmetryhlm}. 

In this paper, we take a different approach, constraining the 3 degrees of freedom as much as possible using only the dominant (2,2) mode. The (2,2) mode of hybrids constructed this way will thus coincide with hybrids constructed only for the (2,2) mode, and  two hybrids constructed with different sets of higher modes will exactly coincide on their common modes, which facilitates comparisons and studies of the contribution of some specific mode.
In practice, our procedure is as follows. Just as in the single mode case, we parametrize how early (or late) in the evolution we perform our hybridization using a ``hybridization frequency'' $\omega_0$, which defines the ``hybridization time'' $t_0$ through
\be
\frac{d\phi_{2,2}^{\rm PN}}{dt}(t_0)=\omega_0,
\ee
and the length of the time-window over which the waveforms are aligned in time as $\Delta t$. Considerations on how to choose these two parameters have been described in the previous section. We can then first determine $\tau$ by minimizing the same quantity as in the single mode case,
\begin{equation}
\Delta(\tau; t_0,\Delta t)=\int_{t_0}^{t_0+\Delta t}
     \left( \omega^{PN}_{2,2}(t) - \omega^{NR}_{2,2}(t-\tau)\right)^2 dt,
\label{eq:omegaintegral22}
\end{equation}
since only the frequencies enter in $\Delta(\tau; t_0,\Delta t)$ so that the determination of $\tau$ decouples from that of the phase offsets $(\psi_0,\varphi_0)$. Before moving on to the determination of  $(\psi_0,\varphi_0)$, let us recall that given a code to generate PN waveforms and an NR code, $\varphi_0$ will depend on choices made to generate each individual waveform whereas $\psi_0$ could in principle be computed once and for all by comparing all the convention choices in both codes. In what follows, we assume that $\psi_0$ can only take the values $0$ or $\pi$ as discussed in Sec.~\ref{sec:notations}. Let us define
\be
\Delta \phi_\lm=\phi_\lm^{\rm NR}(t_0-\tau)-\phi_\lm^{\rm PN}(t_0).
\ee
Then ideally (i.e. assuming that \eqref{generalrelationmodes} holds), we have $\psi_0 + 2 \varphi_0 + \Delta \phi_{2,2}\equiv 0 \bmod{2\pi}$ i.e.
\bea
\varphi_0 \equiv -\frac{\Delta \phi_{2,2}+ \psi_0}{2}  \quad \bmod{\pi}\label{22mode},
%\psi_0 + m \phi_0 + \Delta \phi_\lm &\equiv 0 \bmod{2\pi}\label{33mode}
\eea
which gives 2 solutions for $\varphi_0$ in the interval $[0,2\pi[$ if $\psi_0$ is previously known and 4 solutions if $\psi_0$ is unknown but restricted to  $\psi_0 \in \{0,\pi\}$:
\be
\label{eq:phi0psi0explicitexpr}
(\psi_0,\varphi_0)=\left(\kappa \pi, -\frac{\Delta \phi_{2,2}}{2}+\left(\kappa'-\frac{\kappa}{2}\right) \pi \,\bmod{2\pi}\right)
\ee
with $\kappa\in\{0,1\}$ and $\kappa'\in\{0,1\}$. To lift this degeneracy, we need information from at least one of the higher modes, say $(\ell_*,m_*)$, and we use the one with the largest amplitude, typically the (3,3) mode unless it is zero for symmetry reasons. If again both waveforms were infinitely accurate, \eqref{generalrelationmodes} would imply
\be
\psi_0 + m_* \varphi_0 + \Delta \phi_{\ell_*m_*} \equiv 0 \bmod{2\pi},
\ee
but in the presence of waveform errors this will not hold for any of our four solutions. However, we can choose the solution that is the closest to satisfying this equation, which is uniquely determined only in the case where $m_*$ is odd. Note that in the case where only even $m$ modes are available, $\varphi_0$ needs in fact only to be determined modulo $\pi$ since only the combination $m\varphi_0$ appears in the hybrid construction and the two solutions can be discriminated using any even higher mode.

We find that there is a relative $\psi_0$ shift of $\pi$ between the BAM and SXS waveforms, and also between BAM and the 
conventions used in the PN context by Arun et al \cite{Arun:2009mc} and Blanchet \cite{Blanchet2014}.

\begin{figure}[htbp]
\includegraphics[width=\columnwidth]{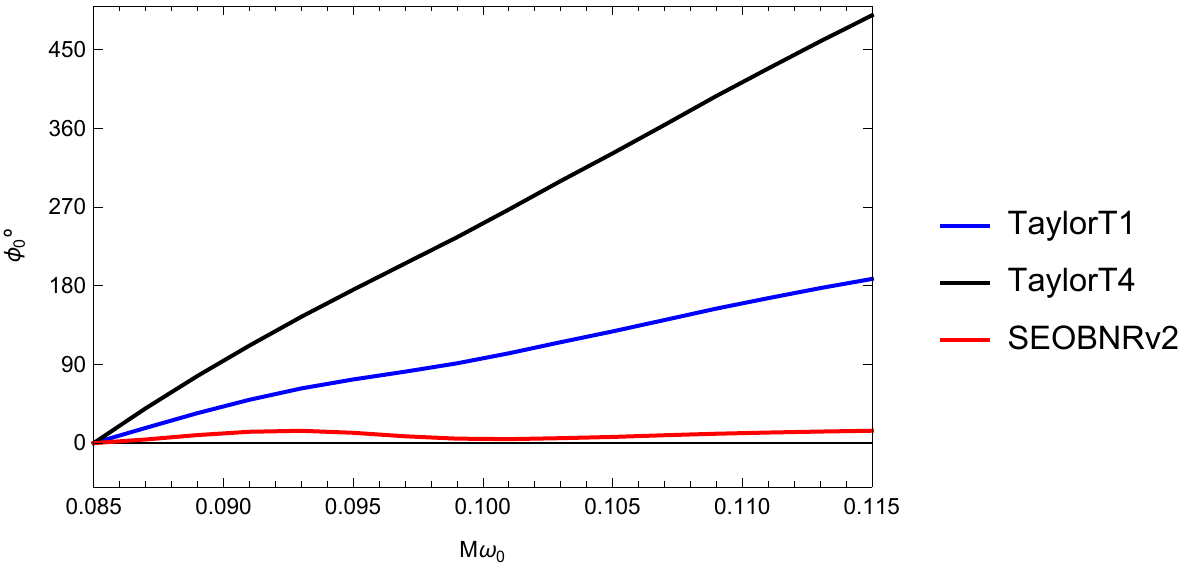}
\caption{Estimations for $\varphi_0$ as a function of $M\omega_0$ for the case of $q=18$ non-spinning BAM NR data hybridized to PN  T4, T1 and SEOBNRv2 (from top to bottom). For T1, the estimation of $\varphi_0$ changes by $\sim \pi$ over the frequency range shown here (corresponding to  $\sim7$ cycles in $h_{2,2}$).
}
\label{fig:psi0q18NS}
\end{figure}

In Fig~\ref{fig:psi0q18NS}, we show the solutions that were found using this procedure on our $q=18$ case hybridized with three different PN approximants as a function of the hybridization frequency. Not surprisingly, the result corresponds to the lower plot of Fig.~\ref{secular22}, which show the dependence of the hybridisation time shift on frequency and PN approximant. In  Fig~\ref{fig:psi0q18NS} we see that both the standard T1 and T4 approximants exhibit large secular trends, indicating a large difference in the orbital phase (or more precisely the phase of the $(2,2)$ modes) between the PN waveforms and the NR result over the frequency range shown here. In contrast (and remarkably), the SEOBNRv2 waveform \cite{Taracchini:2013rva} shows almost no secular trend even though the model was calibrated to NR simulations only up to $q=8$. Evaluating the secular trend of the PN approximant as compared to the NR waveform is an important part of the hybridization procedure. Exactly as in the single mode case, this secular trend translates into some ``hybridization error'' (for instance, the phase of the (2,2) mode for two hybrids built using T1 but with $M\omega_0=0.085$ or $M\omega_0=0.115$ and aligned in the early inspiral will differ at the peak by almost one gravitational wave cycle) but this error has nothing to do with the higher modes themselves and controlling it is not our main focus here. Instead, we will try to identify additional figures of merit for the hybrid that directly quantify the additional error due to the higher modes.

\subsection{Step 2: evaluate residual disagreement between PN and NR at the matching point}

We now investigate the residual phase and amplitude disagreements between PN and NR at the matching point and define appropriate quantities to describe this disagreement, while we postpone the analysis of the main source of this disagreement to Sec.~\ref{sec:errorsources}.

\begin{figure*}[htbp]
%\centering
\includegraphics[width=.9\columnwidth]{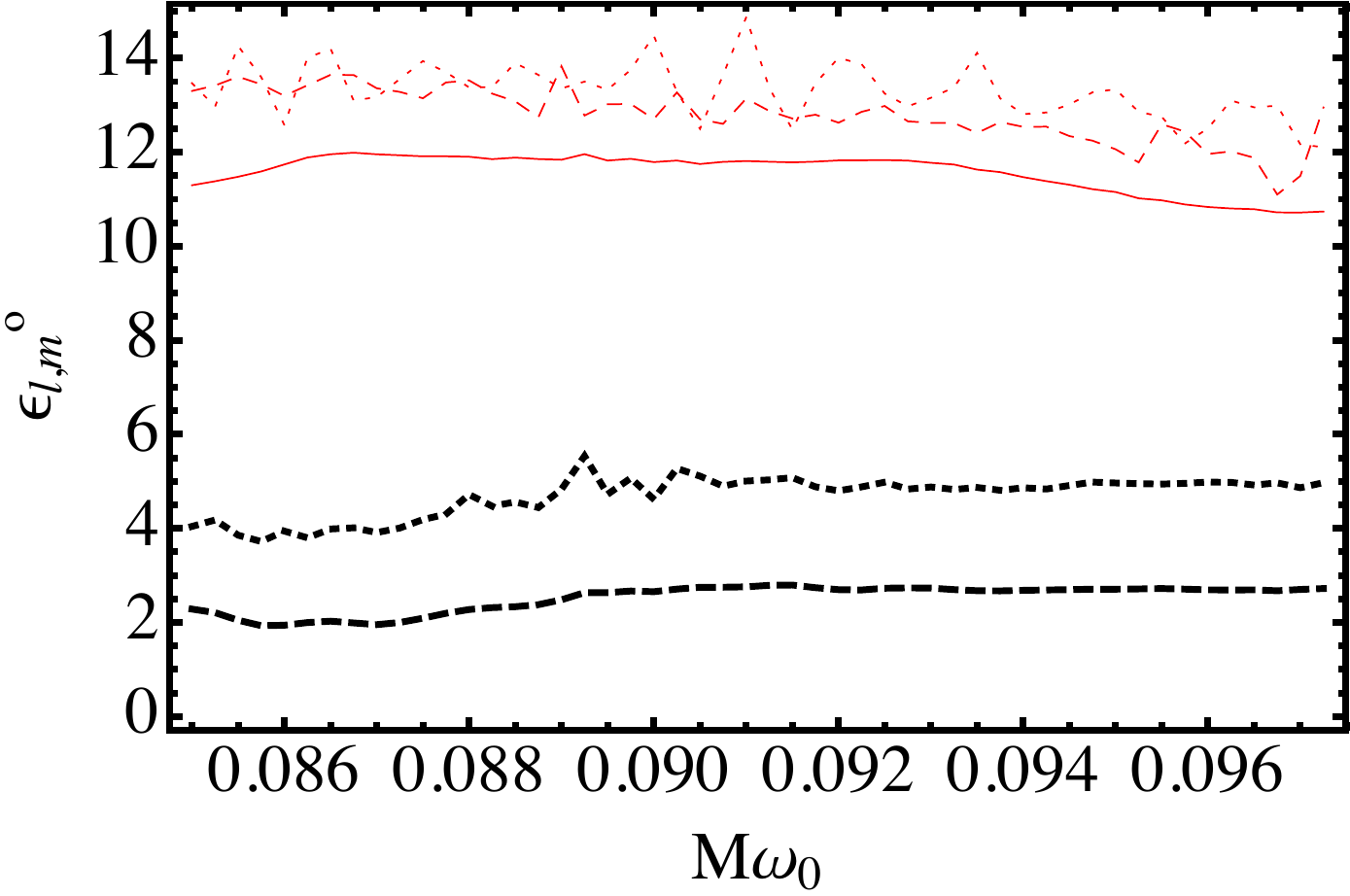}
\includegraphics[width=1.1\columnwidth]{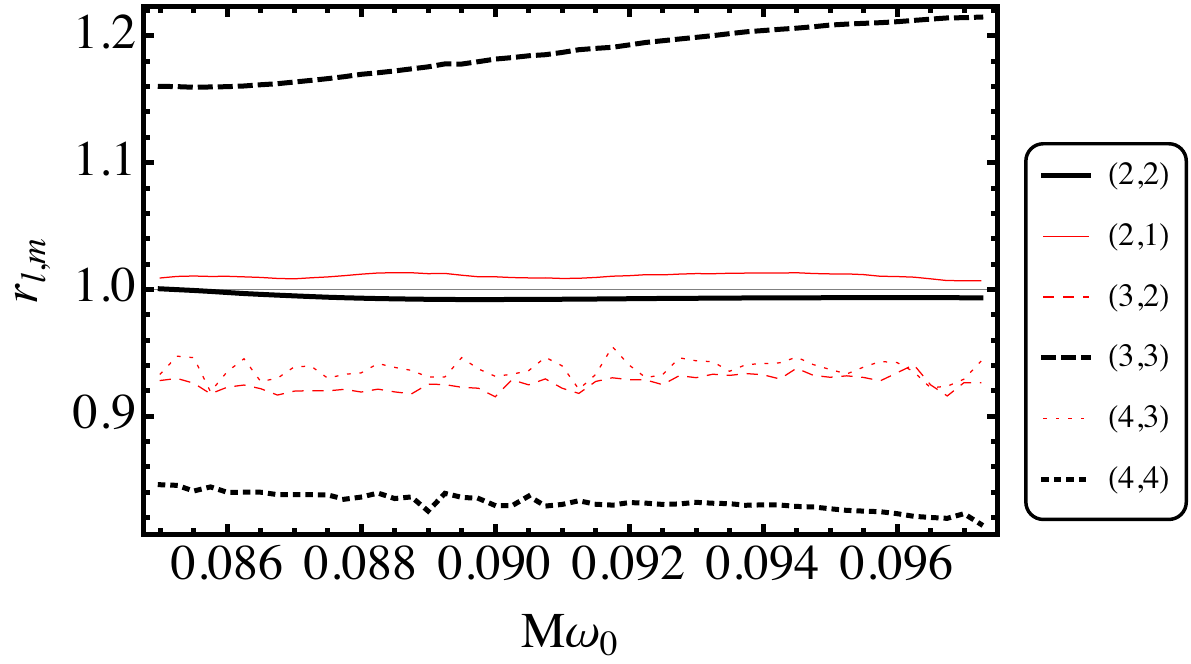}
\caption{Left: Phasing errors $\epsilon_\lm$ (in degrees) for the $q=18$ non-spinning BAM data hybridized with  T1 as a function of $M\omega_0$. Note the lower values for  $\ell=m$ modes (black) as compared to $\ell \neq m$ (red). Right: Amplitude ratios $r_\lm$ for the same hybrid construction. Note how modes with larger $m$, i.e. with larger frequencies, tend to show larger amplitude disagreements.}
 \label{fig:epsilonlmq18NS}
\end{figure*}

In the idealized case, correcting for the differences in conventions using $(\varphi_0,\psi_0)$ is sufficient to ensure that all the phases are continuous between the PN and NR waveforms at the matching point, i.e. that the quantities
\be
\label{eq:defepsilons}
\epsilon_\lm=\Delta \phi_\lm + \psi_0 + m \varphi_0,
\ee
which are functions of the hybridization frequency $\omega_0$ are all zero. In practice, this is not the case and we will use these quantities as measures of the residual phase disagreements. The values for our example $q=18$ case are shown for the most important modes in Fig.~\ref{fig:epsilonlmq18NS} (left) and are typically of a few degrees for $m=\ell$ modes and $~10-15$ degrees for $m=\ell-1$ modes. Apart from the values themselves, one important feature is the fact that unlike for $\varphi_0$, these remain roughly constant over the range of hybridization frequencies explored here. This is a consequence of the fact that we are essentially measuring phase differences between the higher modes \emph{after aligning the  (2,2) modes at} $\omega_0$ via the term $m\varphi_0$ in Eq.~\eqref{eq:defepsilons}, which effectively absorbs the secular dephasing shown in Fig.~\ref{fig:psi0q18NS}. In other words, the $\epsilon_{\ell,m}$ really quantify the residual differences between PN and NR in the dephasings between the higher modes and the (2,2) mode, factoring out the error in tracking the orbital phase (or equivalently that of the (2,2) mode) of the system.
Regarding the amplitude discrepancies, we can simply define the ratio
\be
\label{eq:defrlm}
r_{\ell m}=\frac{|h^{\rm NR}_{\ell m}(t_0-\tau)|}{|h^{\rm PN}_{\ell m}(t_0)|}
\ee
which is plotted in Fig.~\ref{fig:epsilonlmq18NS}  (right). We find that modes with higher frequencies show larger amplitude disagreements. We will perform a detailed analysis of the phase and amplitude errors in Sec.~\ref{sec:errorsources}. 

\subsection{Step 3: hybrid construction}

\begin{figure*}[tpb]
%\centering
\includegraphics[width=\columnwidth]{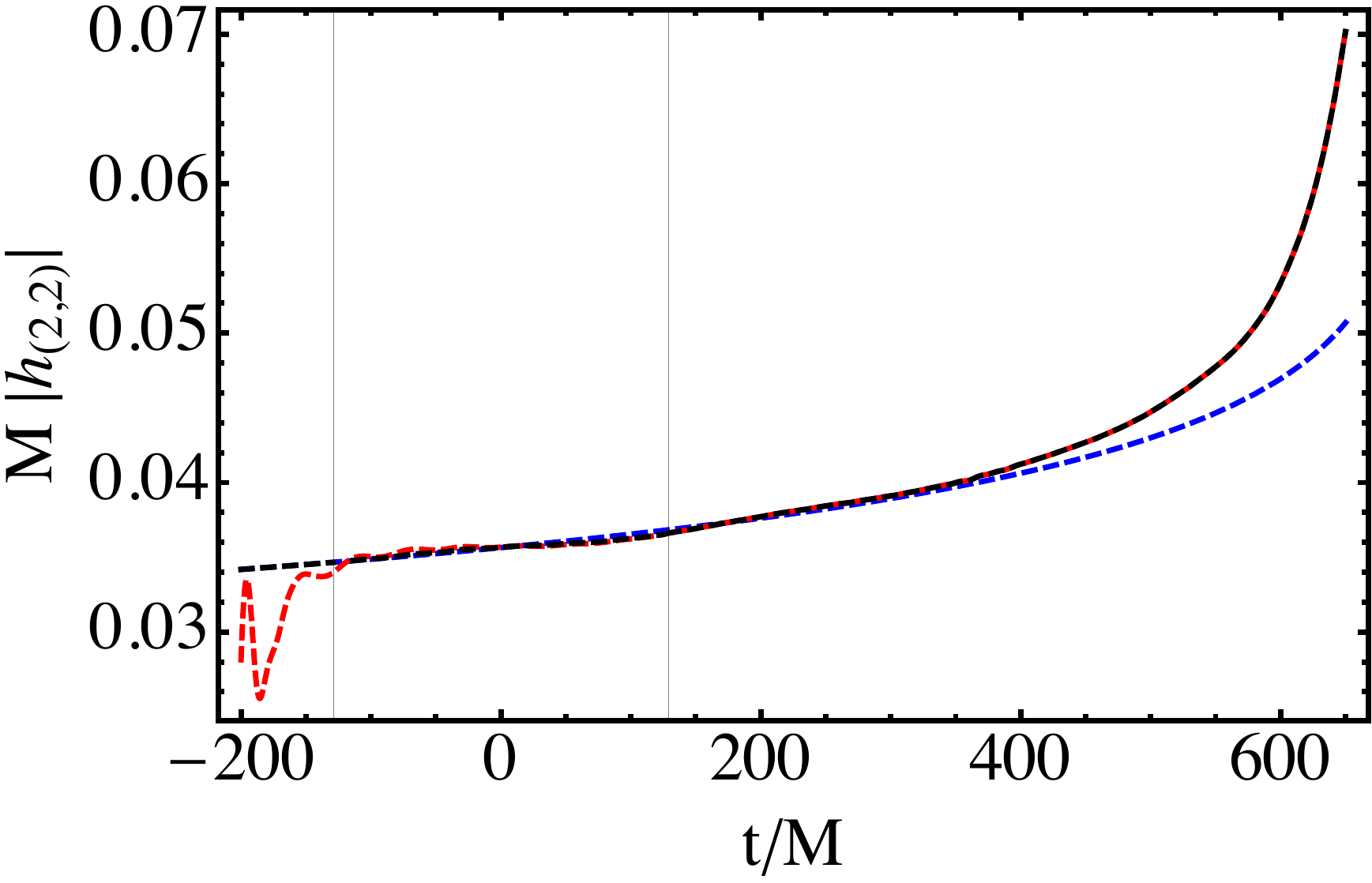}
\includegraphics[width=\columnwidth]{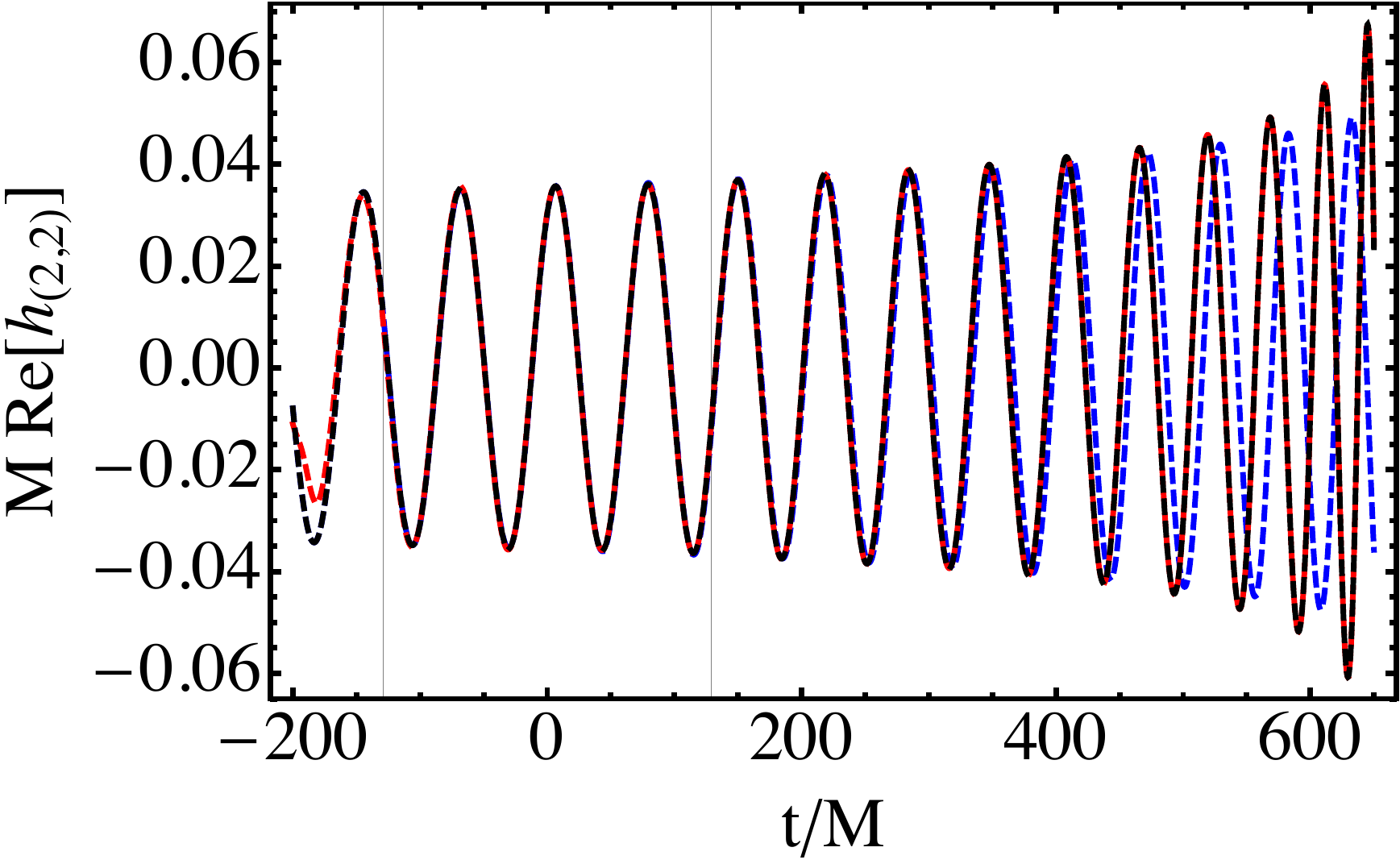}
\includegraphics[width=\columnwidth]{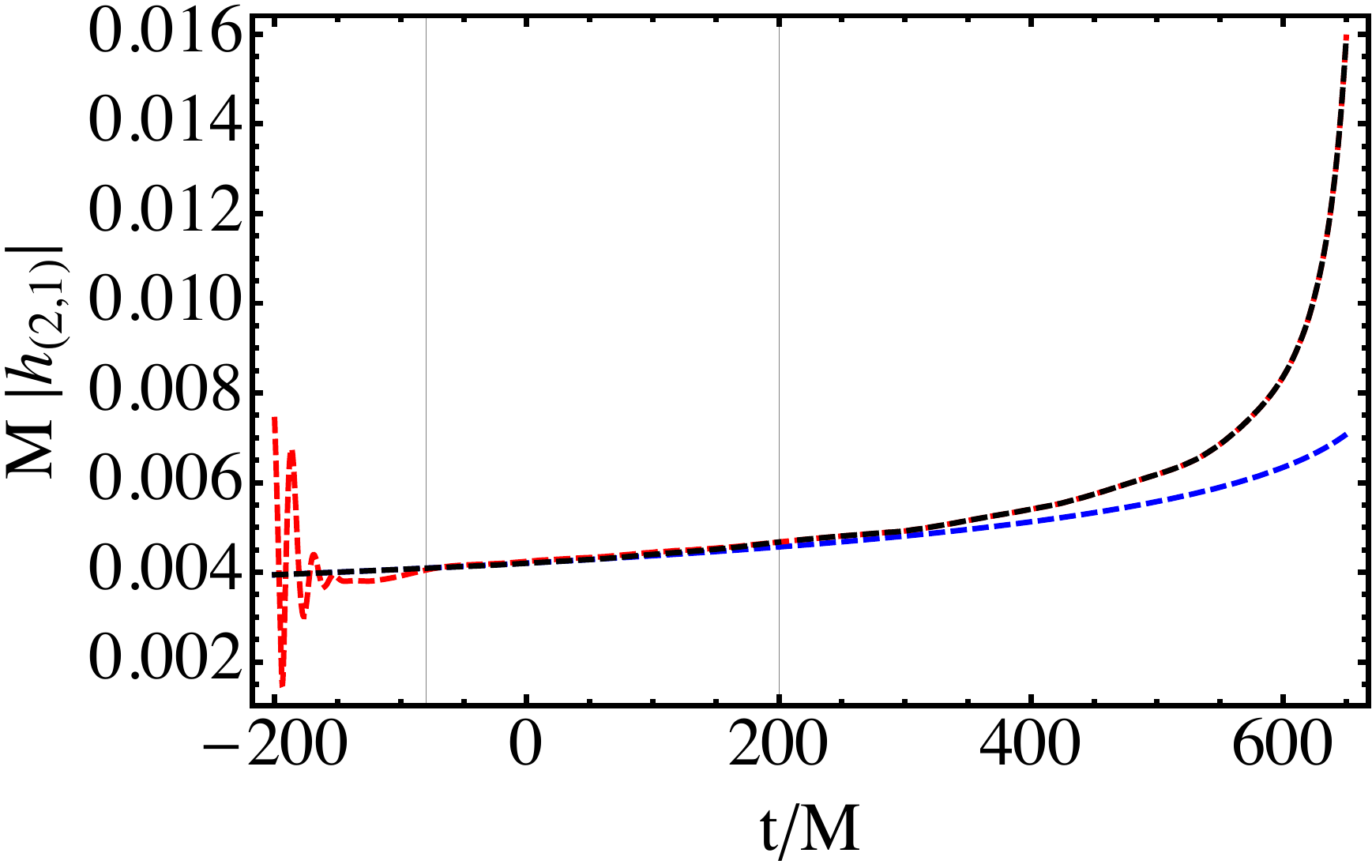}
\includegraphics[width=\columnwidth]{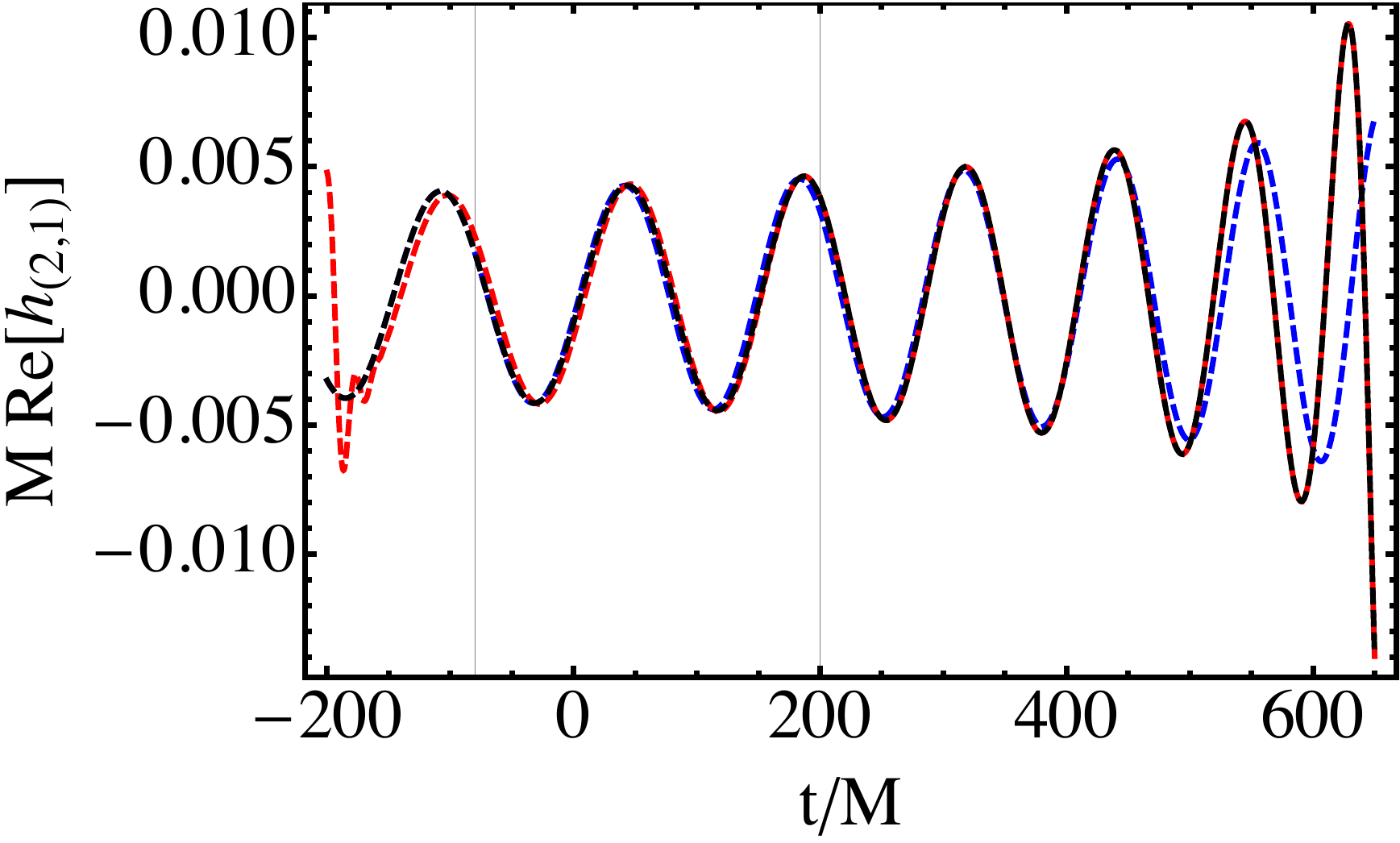}
\includegraphics[width=\columnwidth]{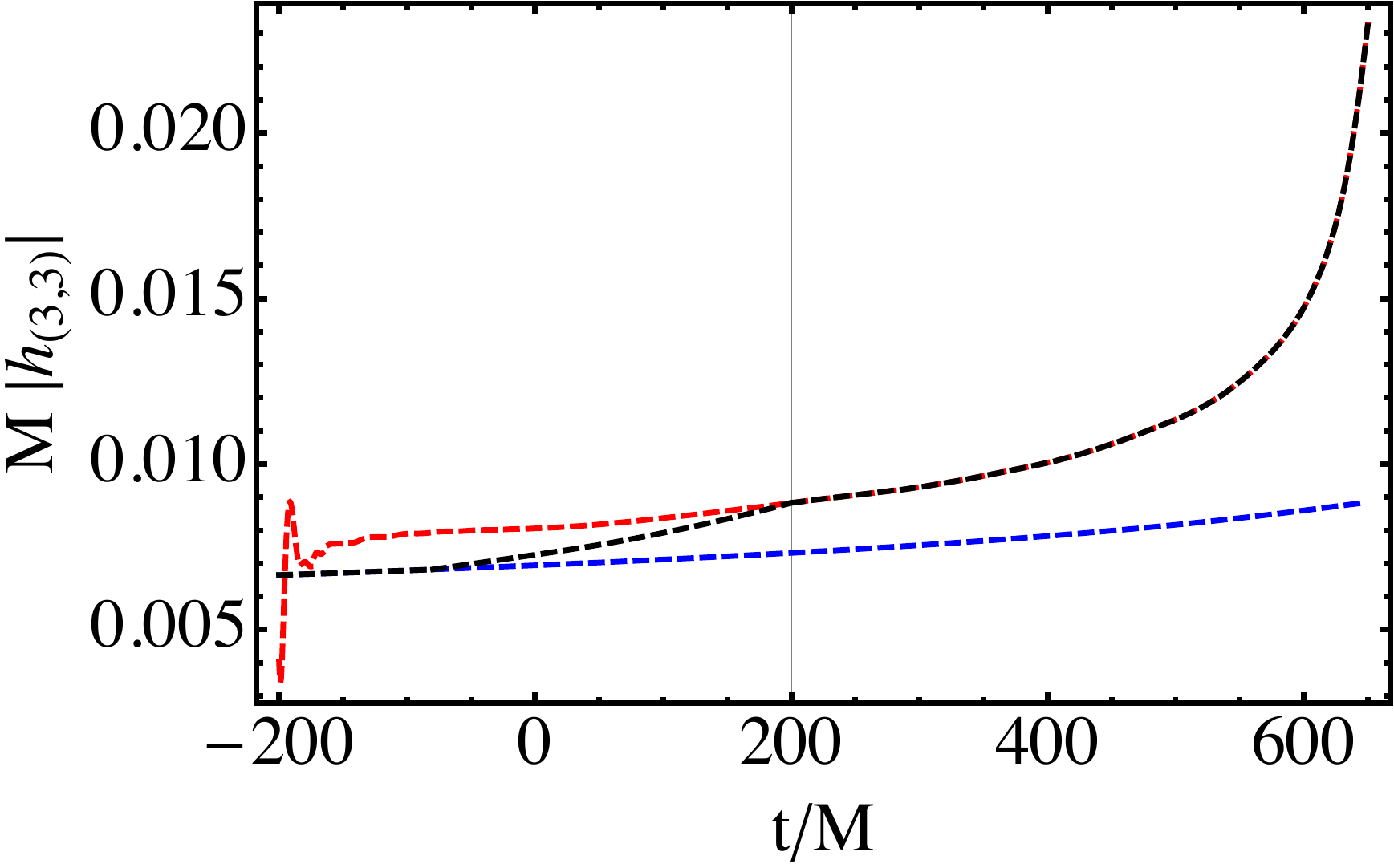}
\includegraphics[width=\columnwidth]{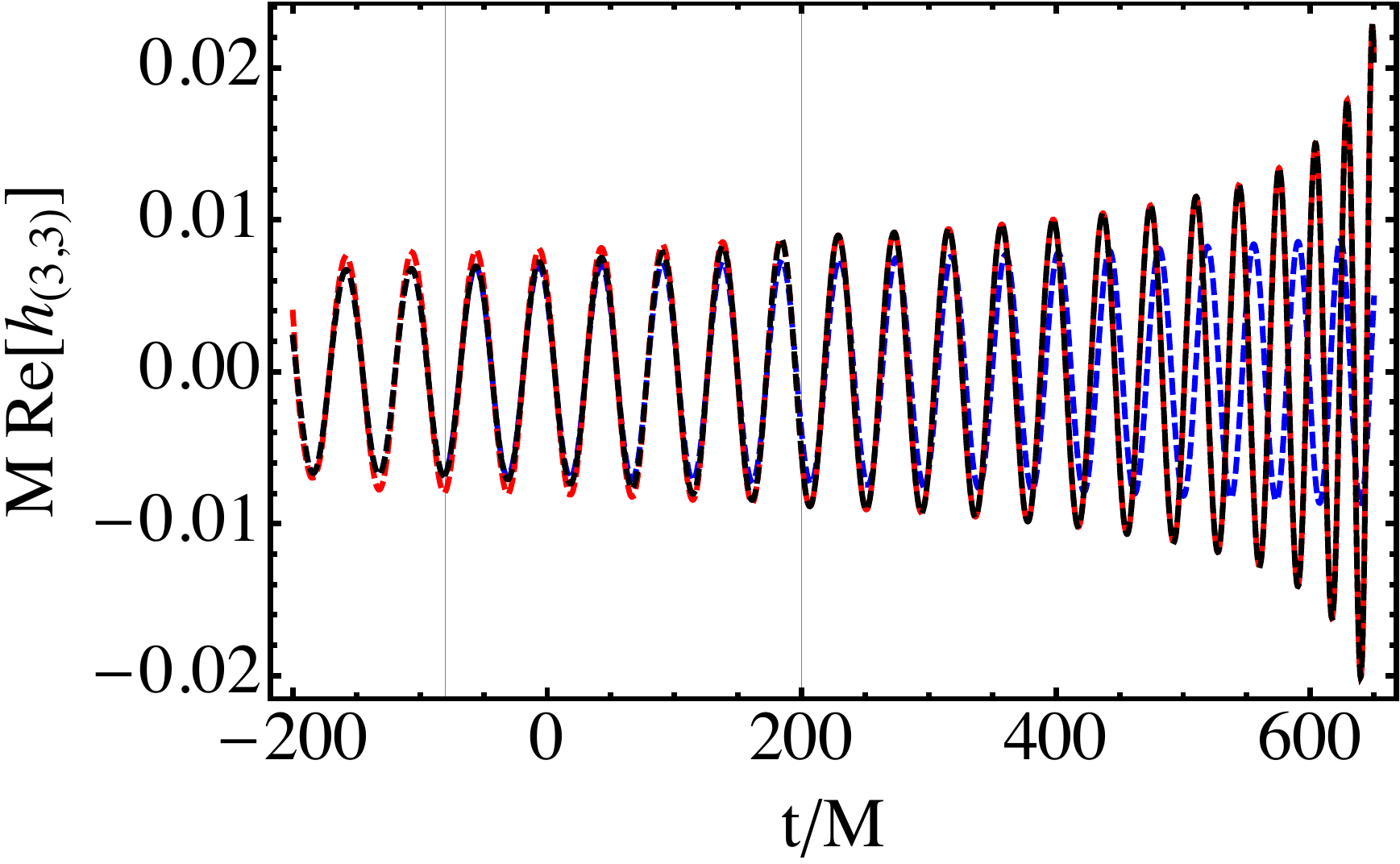}
\caption{Amplitude (left) and real part (right) of non-spinning BAM $q=18$  $(2,2)$, $(2,1)$ and $(3,3)$ modes
hybridised with  T1, from top to bottom. We show PN (blue), NR (red) and hybrid (black) modes and focus on the hybridization region. The $(2,1)$ mode is a typical example of good amplitude agreement and large $\epsilon_\lm$, while the $(3,3)$ mode is a typical example of small $\epsilon_\lm$ and poor amplitude agreement. Note the usage of different blending windows from the one used for the $(2,2)$ mode.}
\label{fig:hybridamplitudesq18lequalm1}
\end{figure*}

We can finally proceed to constructing the higher hybrid modes as piecewise functions in a similar fashion as in Eq.~\eqref{eq:hybrid22}, however we now allow different blending functions for different modes to deal with noisier low amplitude modes. With the notation $[t_0^{\ell,m}-\tau,t_0^{\ell,m}-\tau+\Delta t^{\ell,m}]$ for the interval used for each mode and $w^\pm_{\ell,m}(t)\equiv w^\pm_{[t_0^{\ell,m}-\tau,t_0^{\ell,m}-\tau+\Delta t^{\ell,m}]}(t)$ the associated blending functions, we define
\begin{widetext}
\begin{eqnarray}
\begin{aligned}
 h_\lm(t)=\left\{
 \begin{array}{ll}
e^{i(m\varphi_0+\psi_0)}  h^{PN}(t+\tau) &  t<t_0^\lm-\tau\\
 w_{\ell,m}^-(t)e^{i(m\varphi_0+\psi_0)} h^{PN}(t+\tau) + w_{\ell,m}^{+}(t)  h^{NR}(t) &  t_0^{\ell,m}-\tau<t<t_0^{\ell,m}-\tau+\Delta t^{\ell,m}\\
h^{NR}(t) & t_0^{\ell,m}-\tau+\Delta t^{\ell,m}<t
 \end{array}
  \right .
 \end{aligned}
\label{eq:hom}
\end{eqnarray}
\end{widetext}
with $t_0 \in (t_0^\lm , t_0^\lm+\Delta t)$. Fig.~\ref{fig:hybridamplitudesq18lequalm1} shows three of the resulting hybrid modes.

\section{Hybrid higher modes: phasing and amplitude errors}
\label{sec:errorsources}

In the previous section, we introduced figures of merit to quantify the residual disagreements between the PN and NR higher modes both for the phase (see Eq.~\eqref{eq:defepsilons}) and for the amplitude (see Eq.~\eqref{eq:defrlm}), after correcting for the differences in conventions by aligning the (2,2) modes.
We devote the present section to identifying the main source of these disagreements among the errors affecting both computations and described in Sec.~\ref{sec:sourcesoferror}. In particular, we find that the phasing errors $\epsilon_{\ell m}$ are dominated by the fact that the waves are extracted at finite radius in the NR simulations. Regarding the amplitude errors $r_{\ell m}$, we observe that extraction radius in NR simulations plays an important role (dominant for some modes), but that PN truncation errors are dominant for some other modes. For a detailed analysis of errors in higher modes for a simulation of non-spinning $q=4$ numerical simulations see \cite{Sperhake:2010tu}. One of the main conclusions there is that extrapolation of gravitational waveforms to infinite extraction radius is particularly important for subdominant multipoles with $\ell\neq \vert m\vert$.

For this study, we have looked at a variety of physical configurations (mass ratios, spins), and we focus on those configurations for which several NR simulations have been performed using different codes (and therefore different numerical setups, gauge conditions and initial data). We illustrate our results with the case of a $q=8$ non-spinning binary simulated using the BAM code  \cite{Bruegmann:2006at,Husa:2007hp}, and also available in the public SXS catalogue \cite{SXS}. This is an interesting case with strong higher mode contributions due to the large mass ratio and where wave signals are available at different resolutions for both codes, as well as several extraction radii. Furthermore, the SXS data sets are also extrapolated to null infinity at different polynomial orders, see \cite{Boyle:2009vi} for a discussion of different methods, and \cite{Taylor:2013zia} for a comparison with characteristic extraction results. Note that the SXS waveform is significantly longer, which allowed us to hybridize at a frequency as low as $M\omega_0=0.043$, when the BAM waveform requires $M\omega_0>0.080$. This means hybridizing $40$ and $9.2$ gravitational wave cycles (in the 22 mode) before merger respectively.

\subsection{Amplitude errors}

\begin{figure*}[htbp]
\includegraphics[width=\columnwidth]{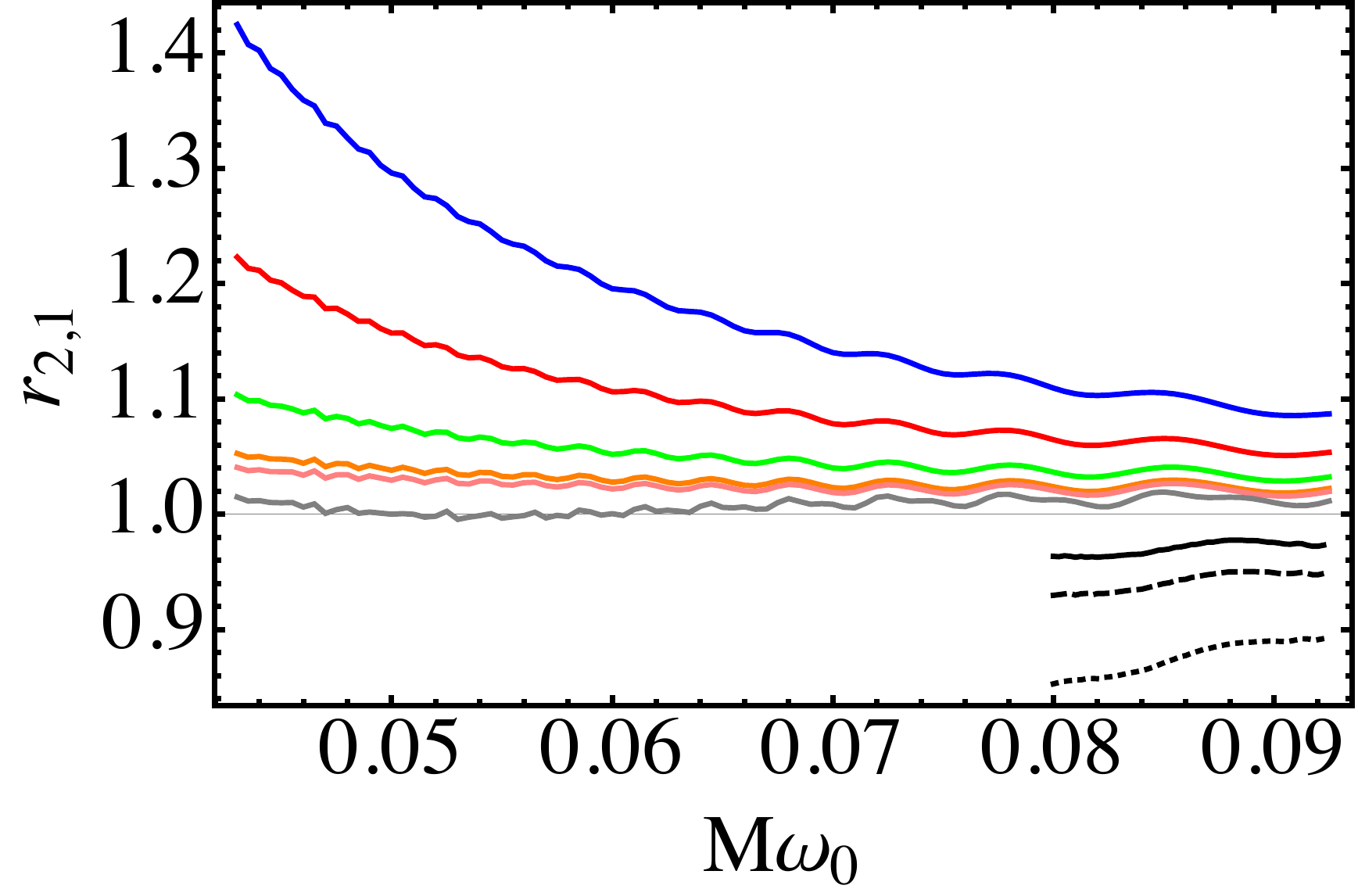}
\includegraphics[width=\columnwidth]{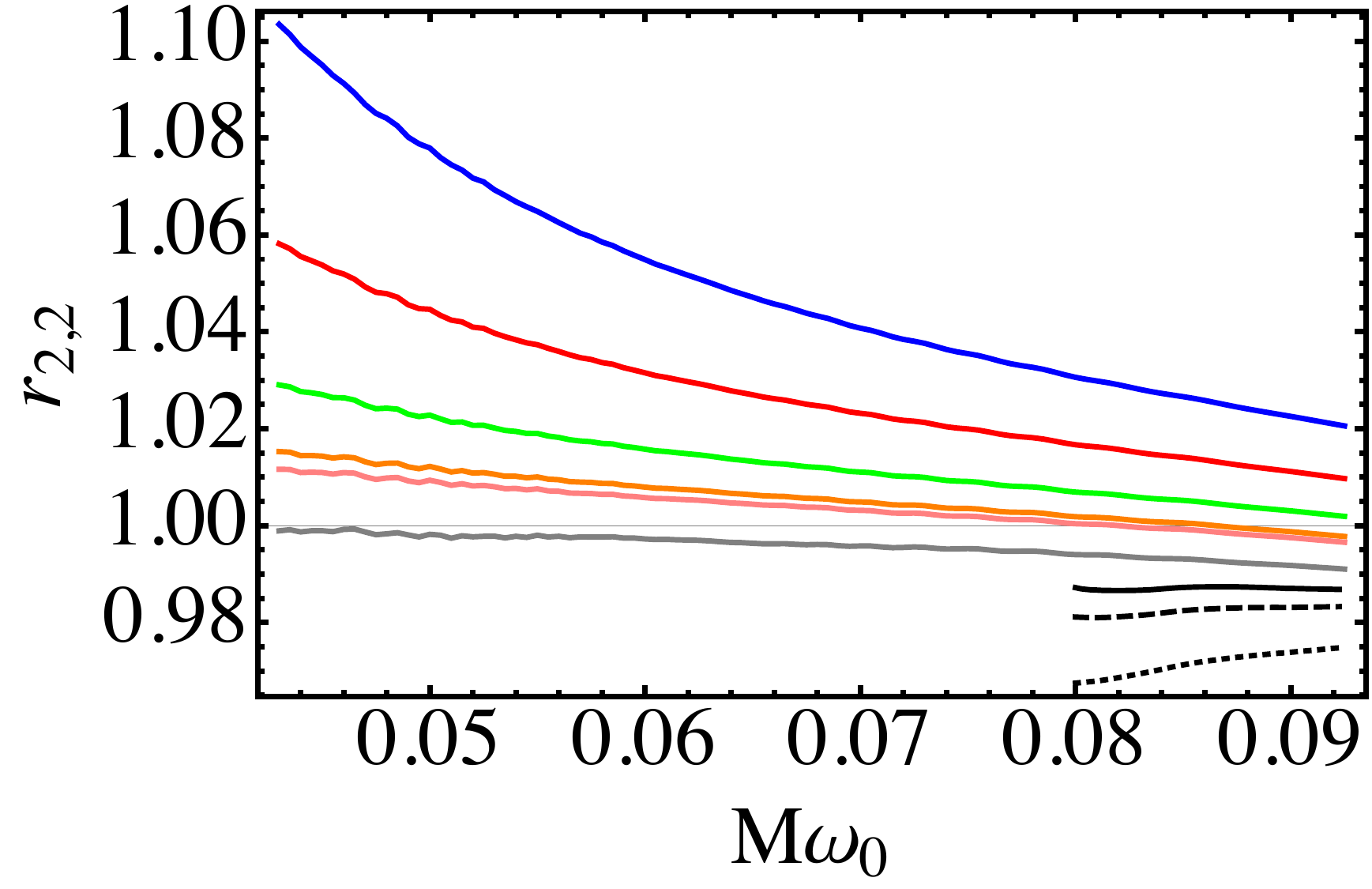}
\includegraphics[width=\columnwidth]{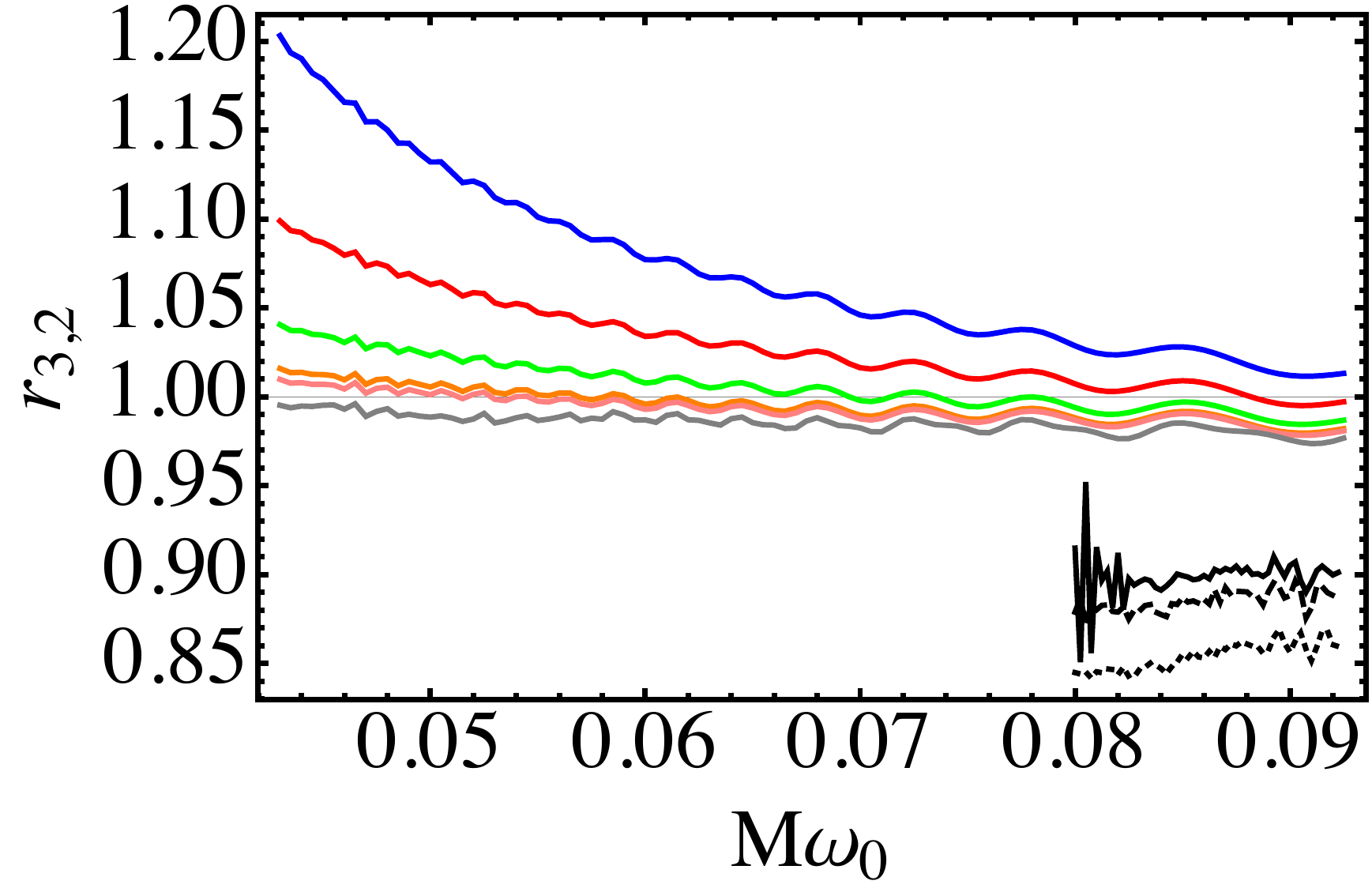}
\includegraphics[width=\columnwidth]{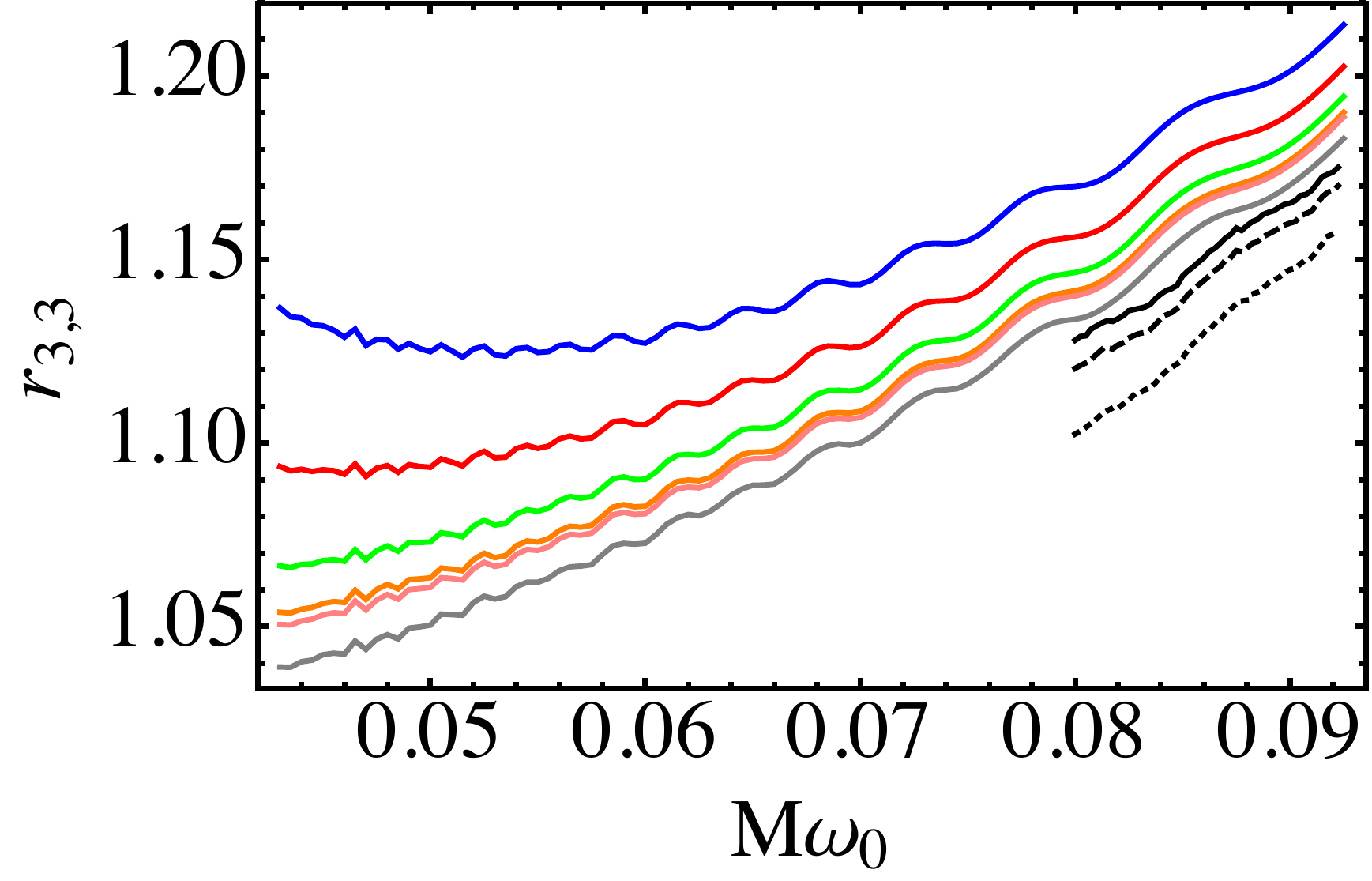}
\caption{Ratio of the NR/PN amplitudes as a function of matching frequency $M\omega_0$ for a $q=8$ non-spinning binary for several NR waveforms matched to T1. The simulations correspond to $r=(100,133,190,266,307)M$ and extrapolated SXS data (in color and downwards) and BAM $r=(60,80,100)M$ (black and upwards).}
\label{fig:Amps}
\end{figure*}

We first focus on the residual amplitude discrepancies measured by the ratios $r_{\ell m}$ defined in Eq.~\eqref{eq:defrlm} and start by investigating the effect of finite radius extraction on these quantities.
Fig.~\ref{fig:Amps} shows the amplitude ratio for different modes as a function of the hybridization frequency $\omega_0$, and for different extraction radii for both BAM and SXS, including the SXS waveform extrapolated to null infinity with a polynomial of order $N=4$. Note that all these curves are amplitude comparisons between NR and PN, but by taking the ratio of two curves, one obtains a direct comparison between two NR results.

While the amplitude of the waves extracted at finite radius typically differs from that of the extrapolated waves by of the order of one percent for the highest radii available, significant errors arise when waves are extracted closer to the source, and errors are quite different for data sets computed with different codes.  With the exception of the $(3,2)$-mode, e.g. $r=100$ BAM data are significantly closer to the extrapolated result than the corresponding SXS $r=100$ curve. This is not surprising, since finite radius errors depend on gauge conditions, and the choice of lapse and shift are indeed different for both codes. We have at present no explanation why the BAM $(3,2)$-mode shows  anomalous behavior as compared to the other modes.
We note that, for any given finite extraction radius, the error becomes larger at lower frequencies. This is the expected consequence of the fact that as the frequency of the waves increases (or equivalently as their wavelength decreases), the wavezone (defined by $r\gg\lambda$) extends to smaller radii.

Having highlighted the effect of extraction radius on the NR amplitude, we now come back to the comparison between NR and PN and restrict our attention to the extrapolated SXS amplitude on the NR side. For the $(2,1)$, $(2,2)$ and $(3,2)$ modes shown in Fig.~\ref{fig:Amps}, the ratio remains almost constant and differs from 1 by at most two percent over the whole range of frequencies considered. On the contrary, the ratio for the $(3,3)$ mode shown in the lower right panel features a strong secular trend, significantly departing from $1$ at high frequencies by a few $10\%$. We observed the same behavior in the other relevant higher modes (i.e. the $(4,3)$, $(4,4)$ and $(5,5)$) not plotted here. Some smaller (but still of the order of several percent) disagreements are visible at low frequencies. The agreement between the different NR curves (extrapolated SXS and the outermost extraction radius for BAM), at least to a much higher degree than the disagreement between PN and NR, and the fact that this discrepancy grows with frequency suggest that the main source of error here is that caused by the PN truncation.

\begin{figure*}[tbp]
\includegraphics[width=\columnwidth]{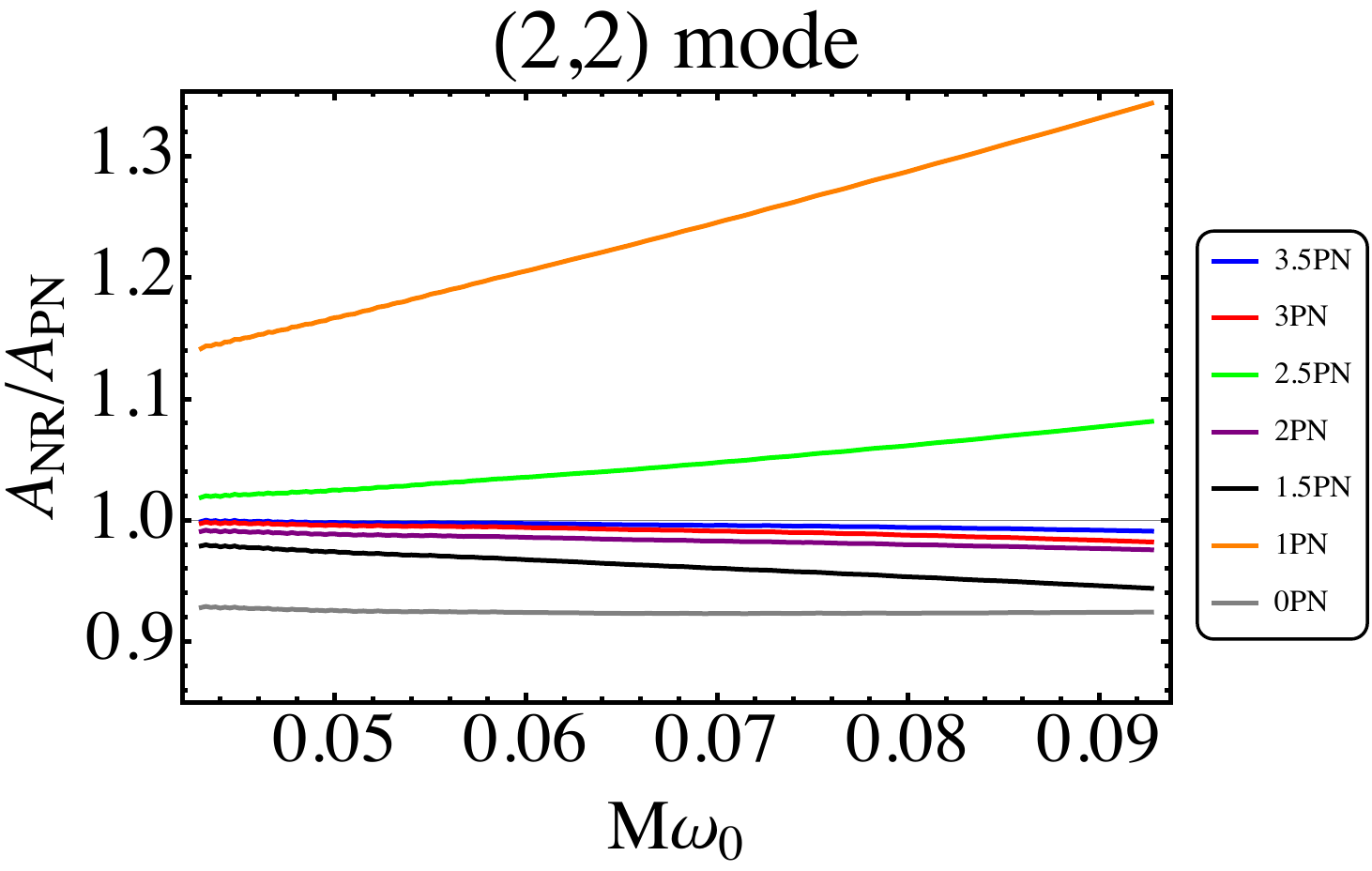}
\includegraphics[width=\columnwidth]{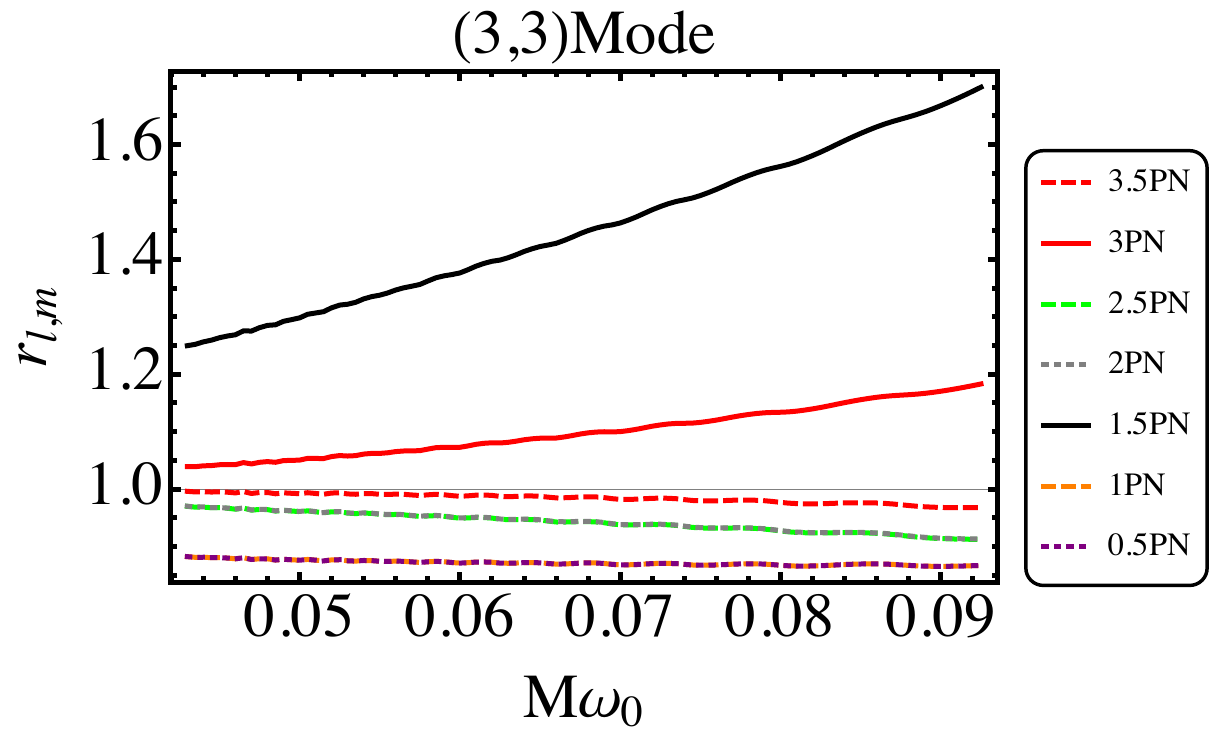}
\caption{Ratio of NR and PN amplitudes of $q=8$ non-spinning SXS NR data extrapolated at $N=4$ matched to  T1 including different PN amplitude corrections as a function of $M\omega_0$. We show the $(2,2)$ (left) and $(3,3)$ (right) modes.}
\label{fig:amptruncationerror}
\end{figure*}

We can analyze this further by looking at how the amplitude ratio varies with the PN order used to compute the modes. Fig.~\ref{fig:amptruncationerror} shows the ratio $r_{\ell m}$ for the $(2,2)$ mode and the $(3,3)$ mode (which exhibited different behaviors in Fig.~\ref{fig:Amps}) computed between the SXS extrapolated NR modes and the PN ones including different PN corrections. We recall here that the $(2,2)$ mode amplitude is known up to 3.5 PN while all the other modes are known up to 3PN. [We note however that while this paper was in preparation, Faye et al. \cite{Faye2015} computed the 3.5PN (non-spinning) contribution to the $(3,3)$ mode. This correction has not been taken into account in the rest of this paper (it would only marginally affect the phasing results shown in the left panel of Fig.~\ref{fig:epsilonlmq18NS} and the right panels of Figures \ref{fig:hybridamplitudesq18lequalm1}, \ref{fig:convPNSXS}, \ref{fig:epsilonsSXS} and \ref{fig:epsilonsSXSq} as well as the match plots in Sec.~\ref{sec:MatchResults}, the qualitative behavior remaining unchanged), and in particular not in Fig.~\ref{fig:Amps} but we do add it in Fig.~\ref{fig:amptruncationerror} where it does have an important effect since it cures the large discrepancy observed in Fig.~\ref{fig:Amps} and brings the agreement between PN and NR to the level of $\sim 2\%$ (the amplitude plots Figs.~\ref{fig:epsilonlmq18NS} (right) and \ref{fig:hybridamplitudesq18lequalm1} (bottom left) would also noticeably improve if this correction was added). This significantly better agreement after improving the PN description of the mode provides additional evidence that PN truncation was the dominant source of error here.]
As usual, studying the PN truncation error is more difficult than studying for instance the convergence with extraction radius because we do not know a priori how the series converges and therefore we cannot extrapolate it. In particular, 
the convergence is not necessarily monotonic and
consecutive corrections can have very different magnitudes (in part because they originate from or mix several physical effects) so the fact that adding a given correction barely changes the result gives no guarantee that the next one will also be negligible. In other words, one cannot estimate the truncation error by comparing the result at $n$ PN and the one at $(n+1/2)$ PN as trivially illustrated by the fact that the 2PN and 2.5PN curves of the right panel of Fig.~\ref{fig:amptruncationerror} are superimposed but differ significantly from the 3PN curve. Despite these caveats, the spread between several consecutive curves gives a sense of how much PN truncation affects the result and the comparison between the left and right panels suggests that this error is significantly larger for the $(3,3)$ mode than for the $(2,2)$ mode where moreover the highest PN orders agree well with NR. We observed a similar spread for the $(4,3)$, $(4,4)$ and $(5,5)$ modes while the $(2,1)$ and $(3,2)$ modes exhibited a behavior similar to the $(2,2)$ mode. Note that while all higher modes are known up to absolute 3PN order (with the convention that the leading order of the $(2,2)$ amplitude is Newtonian), their leading order is $(\ell-2)/2$ PN for even values of $\ell+m$ and $(\ell-1)/2$ PN for odd values of $\ell+m$ so the number of relative corrections actually known for each mode of course varies. However, we highlight that the differences that we observe are not a consequence of a relative higher order knowledge for some modes: both the $(2,1)$ and the $(3,3)$ modes for instance are used with $2.5$ PN relative precision and show a very different behavior. The lack of apparent error systematics regarding PN orders or modes is not untypical for PN results, 
where contributions at different orders often come from very different physical effects, thus their magnitude is hard to anticipate.

\subsection{Comparison of PN and NR phases}
\label{sec:phase_PN_NR}

We now move to identifying the origin of the residual phasing errors $\epsilon_{\ell m}$ defined in Eq.~\eqref{eq:defepsilons}. As discussed in Sec.~\ref{sec:hybridhighermodes}, these quantify the discrepancies between PN and NR in the phase difference between the $(\ell,m)$ mode and the $(2,2)$ mode, i.e. dephasing errors in addition to the difference in tracking the orbital phase of the system. Indeed, we can rewrite Eq.~\eqref{eq:defepsilons} using Eq.~\eqref{eq:phi0psi0explicitexpr} to obtain
\be
\label{eq:epsilonasLambda}
\epsilon_{\ell,m}(\omega_0)=\left(\phi_{\ell m}^{\rm NR}-\frac{m}{2}\phi_{22}^{\rm NR}\right)-\left(\phi_{\ell m}^{\rm PN}-\frac{m}{2}\phi_{22}^{\rm PN}\right)
\ee
where the phases have to be taken at the matching point corresponding to the hybridization frequency $\omega_0$. Note that in principle the rhs of Eq.~\eqref{eq:epsilonasLambda} should also contain an additional term $(2 \kappa' m + (2-m) \kappa)\pi/2)$ originating from convention differences which we assume for simplicity have been reabsorbed in the definition of one of the two waveforms. In other words, we adjust the conventions (but only by integer factors of $\pi/2$ for $\varphi_0$ and of $\pi$ for $\psi_0$) of say the PN waveform so that the $\epsilon_{\ell m}$ vanish in the limit where both the NR and the PN waveform would be infinitely accurate.

The $\epsilon_{\ell,m}$ are more intricate quantities to study than the $r_{\ell m}$ since they not only involve PN and NR but also two different modes. As a first step, we find it useful to focus on the simpler quantities
\be
\label{eq:Lambda}
\Lambda_\lm^{X}(t)=\phi_\lm^{X}(t)-\frac{m}{2}\phi_{2,2}^X(t),
\ee
where X is NR or PN, which only involve either the NR or PN waveform. Note that $\Lambda_\lm^{X}$ is insensitive to a redefinition of the angle $\phi$, i.e.~with our previous notation, to a change in $\varphi_0$. However, a redefinition $\psi_0\rightarrow \psi_0 + \delta \psi_0$ affects \eqref{eq:Lambda} as $\Lambda_\lm^{X}\rightarrow \Lambda_\lm^{X} + \delta\psi_0(1-m/2)$. In particular, we define these quantities with the idea in mind that during the inspiral they should vanish as the frequency decreases, which implies some particular convention for $\psi_0$. In the rest of this section, in order to simplify the discussion, we assume that all the waveforms adopt this convention. Note that the conventions adopted in Arun et al \cite{Arun:2009mc} and Blanchet \cite{Blanchet2014} differ from this by a shift of $\pi$ in $\psi_0$.

In PN, one finds that the  $\Lambda_\lm^{\rm PN}$ are small but non-zero during the inspiral, and vanish in the limit of infinite separation. Consequently, the PN phase of the $(\ell,m)$ mode approximately follows the rule $\phi_\lm^{PN}(t)\approx m \phi_{\rm orb}(t)$ where $\phi_{\rm orb}$ is the orbital phase of the system. More precisely, in the absence of precession, the deviations to this expression are due to imaginary coefficients in the mode amplitudes which only appear at high PN orders for the modes we consider (see Eq.~(327) of \cite{Blanchet2014} for the non-spinning mode amplitudes; note that the spin corrections to the mode amplitudes that are currently known, i.e. up to 2PN, contain no such complex correction).
%Of course, the approximate relation $\phi_\lm(t)\approx m \phi_{\rm orb}(t)$ has to break down eventually during the merger, as it is violated during the ringdown.

\begin{figure*}[htbp]
\centering
\includegraphics[width=\columnwidth]{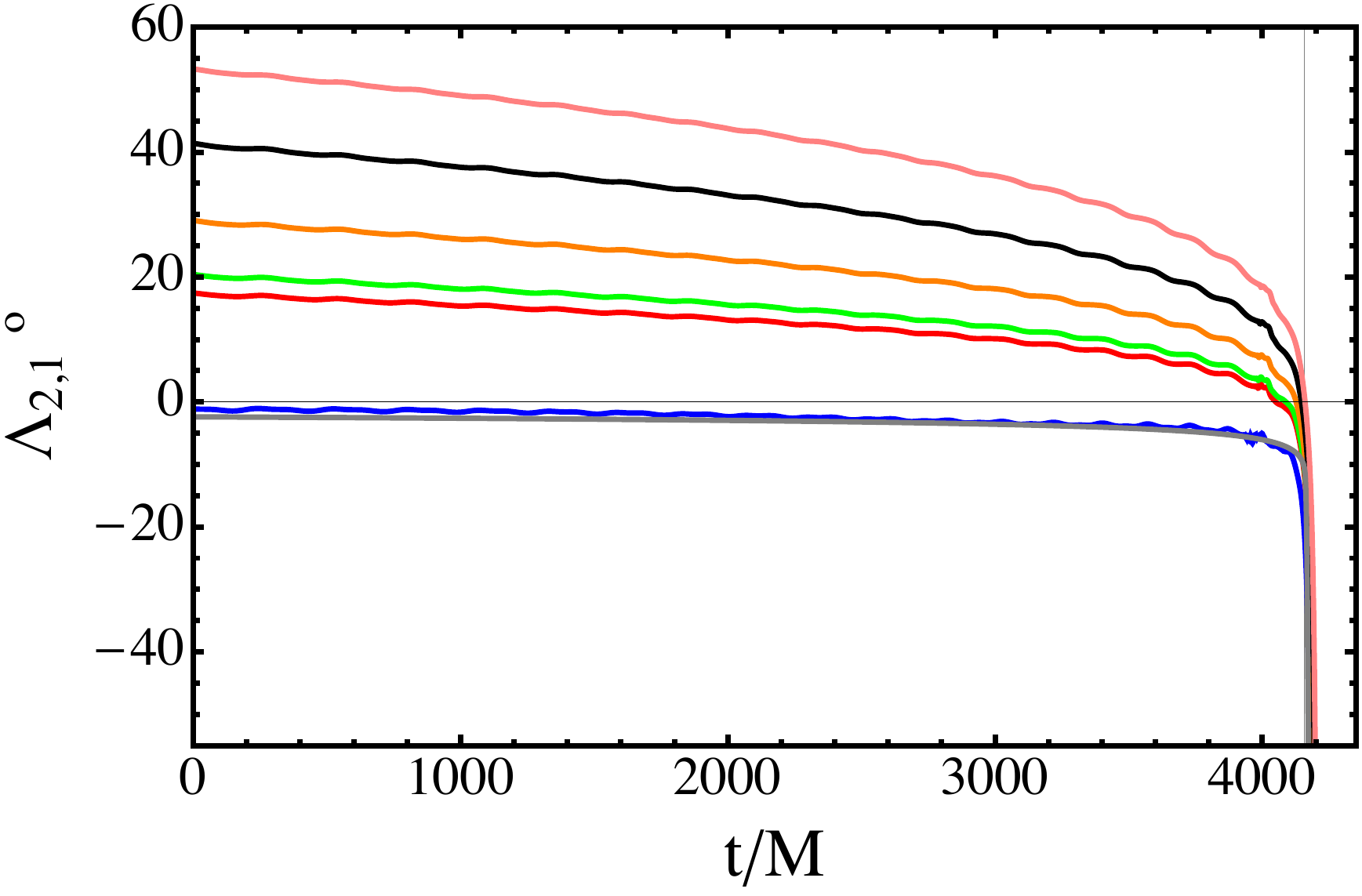}
\includegraphics[width=\columnwidth]{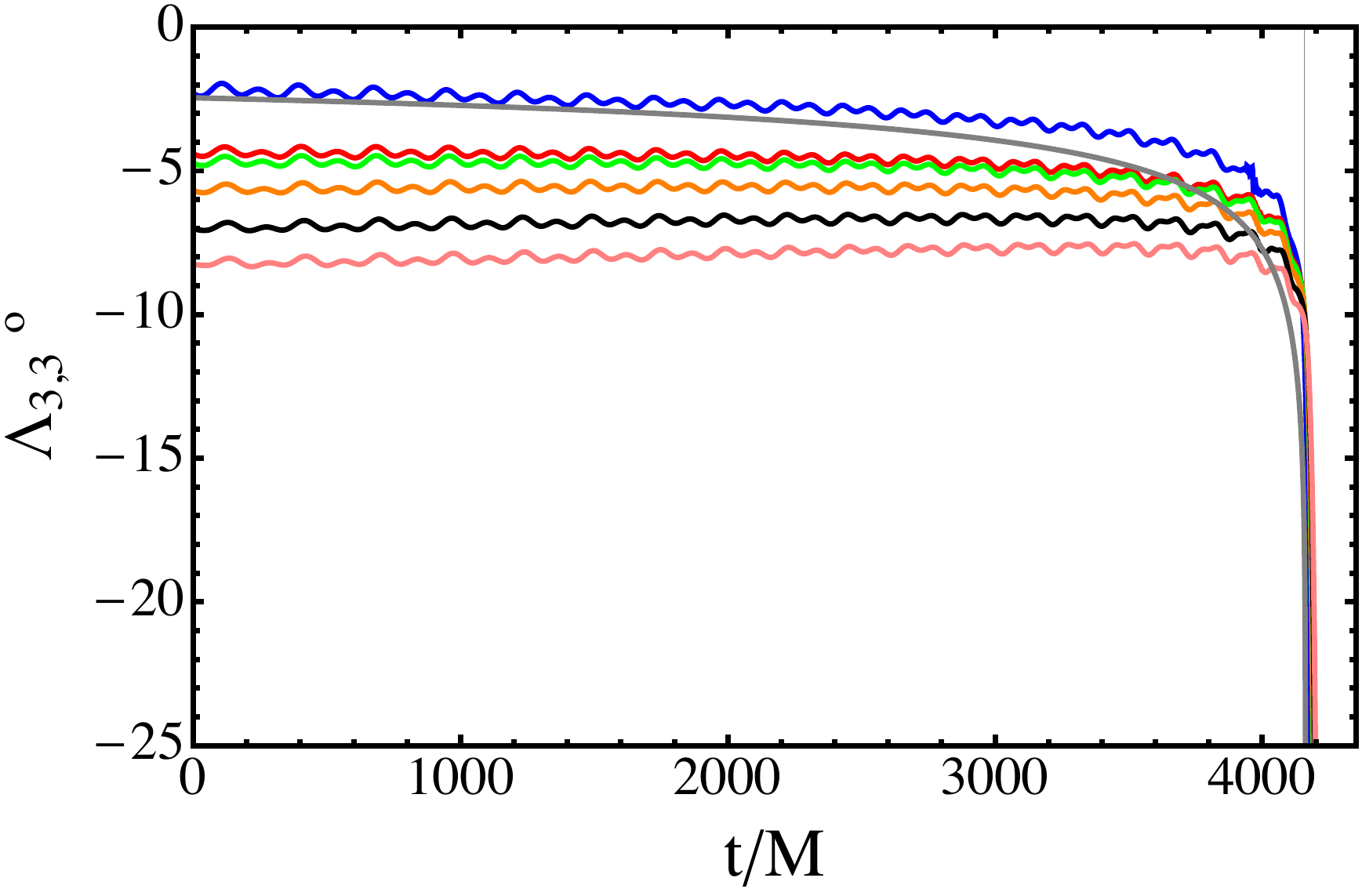}
\caption{Value of $\Lambda_\lm^{\NR} $ for different extraction radii $(100,133,190,266,307)M$ for SXS $q=8$ non-spinning data and PN  T1. We show here the $21$ and $33$ modes. The vertical line denotes the location of the merger. The gray and blue curves correspond to PN and extrapolated $N=4$ data respectively.}
\label{fig:convPNSXS}
\end{figure*}

We find excellent agreement between extrapolated NR and PN regarding the $\Lambda_\lm$, i.e.~in our NR data we find consistent small but non-zero values of $\Lambda_\lm^{\rm NR}$. However, errors resulting from finite extraction radius can be large, and in the following we will discuss the dependence of the quantities $\Lambda_\lm^{\NR} $ on the extraction radius. Since these quantities are functions of time, we first need to align in time waveforms extracted at different radii. To facilitate the comparison with PN, we pick a reference frequency $\omega_0$ early in the waveform and align all the 22 modes with say the one with the largest extraction radius using the same procedure as described in Sec.~\ref{sec:hybrid22}. Fig.~\ref{fig:convPNSXS} shows the (aligned in time) $\Lambda_\lm$ for the $(3,3)$ and $(2,1)$ modes for the $q=8$ non-spinning waveform from the SXS catalog for different extraction radii as well as for the $N=4$ extrapolated numerical waveform and the PN T1 waveform. The alignment has been performed at $M\omega_0=0.043$, and the time coordinate used for the plot has been shifted so that $t=0$ corresponds to $M\omega_0=0.043$ for the extrapolated waveform. For the $(2,1)$ mode, the difference with the extrapolated waveform remains of the order of $15-20$ degrees even for the outermost available radii, and we observe the same behavior for all the $(\ell,\ell-1)$ modes. On the contrary, for the $(3,3)$ mode (and the other $(\ell,\ell)$ modes), the larger finite radii curves only differ from the extrapolated one by of the order of $2$ degrees. Most importantly, in both cases, the extrapolated waveform agrees very well with the PN one, with a typical difference of only one degree, illustrating that the main source of disagreement in the $\Lambda_{\ell m}$ comes from the finite radius extraction.

To quantify this further, we studied the asymptotic behavior of ${\Delta \Lambda}_{\ell m}=|\Lambda^{NR}_{\ell m}-\Lambda^{PN}_{\ell m}|$ (averaged in time over the interval $[1000M,2000M]$ to average out oscillations) as a function of the extraction radius by fitting it to the power law $1^\circ(r/r_0)^n$. The best fit values for $n$ and $r_0$ are shown in Table \ref{tab:convergence} and are consistent with an $1/r$ falloff rate.
\begin{table}[h]
\begin{tabular}{c|c|c|c|c|c}
$(\ell,m)$     &   $(2,1)$         &$(3,2)$         &$(3,3)$         &$(4,3)$         &$(4,4)$\\ \cline{1-6}
$n $          &     -0.967        & -1.015          & -0.941        & -1.038        & -0.947 \\ 
$r_0$      &     3199     & 4215          & 293       &      4182     & 598
\end{tabular}
\caption{%\textcolor{red}{To be rewritten. Also, could we have the value of $r0$? This gives an idea of how far one would have to go to obtain 1 degree} 
Results for $(r_0,n)$ for the fits  ${\Delta \Lambda}_{\ell m}=1^{\circ}(r/r_0)^{n}$ for the case $q=8$ non-spinning SXS NR data matched to  T1. The values suggest an asymptotic  $1/r$-falloff in the inspiral region, where we hybridize. 
%Note how $(\ell,m\neq \ell)$ modes show larger values of $r_0$ indicating a slower convergence.
}
\label{tab:convergence}
\end{table}

\begin{figure*}[htbp]
\centering
\includegraphics[width=\columnwidth]{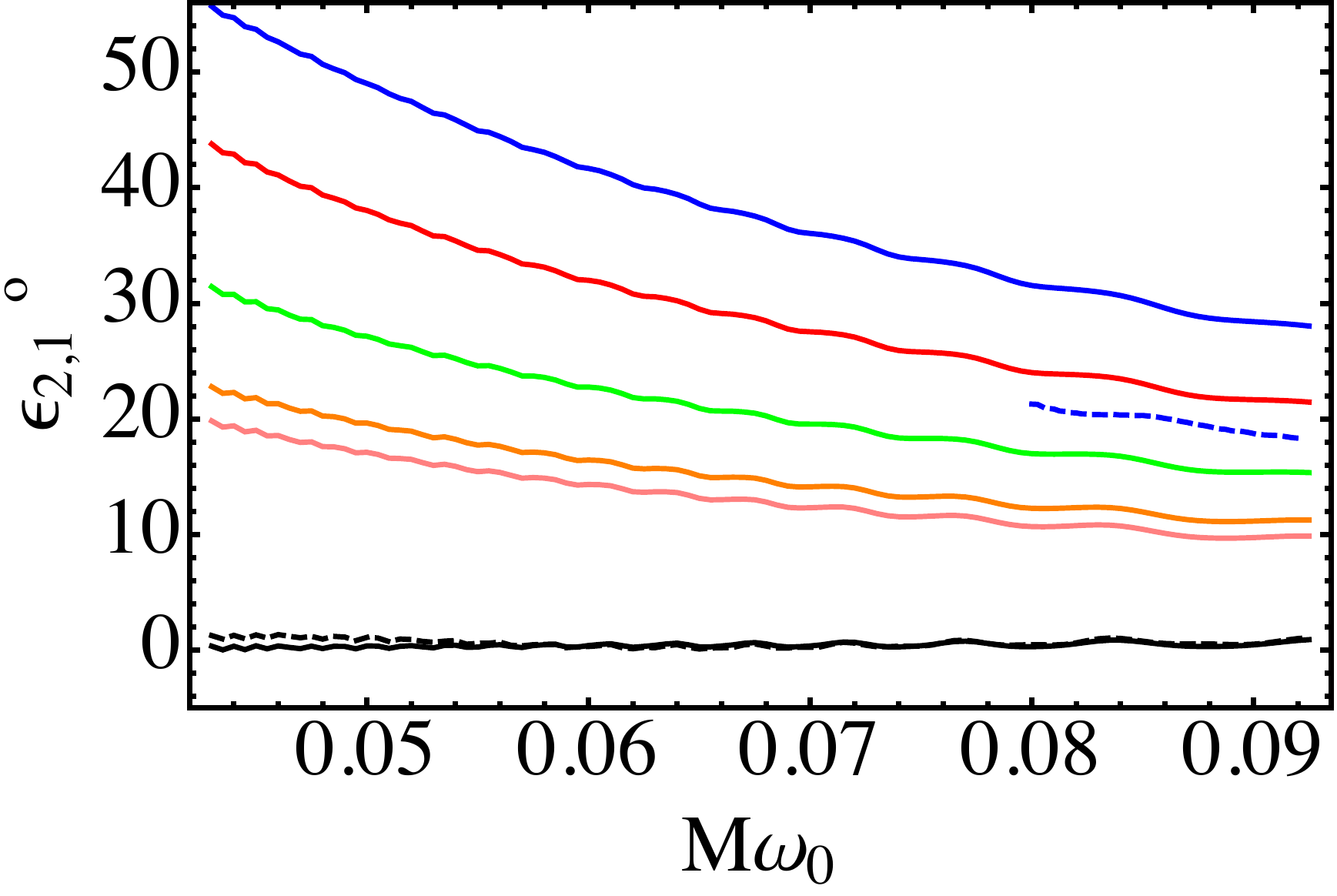}
\includegraphics[width=\columnwidth]{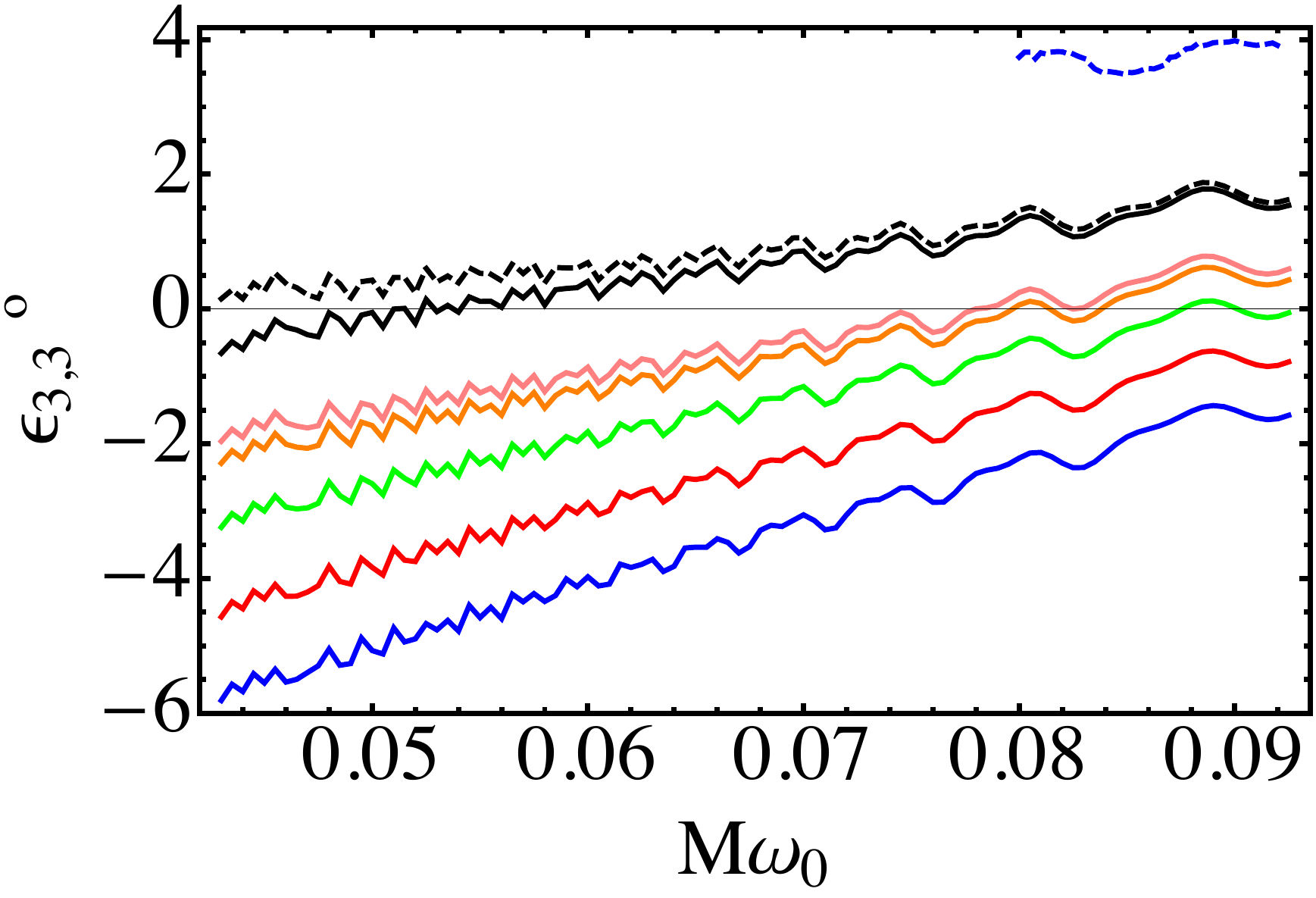}
\caption{$\epsilon_\lm$ values for a the $q=8$ non-spinning system from the SXS catalog for extraction radii $r=(100,133,190,266,307)M$ (in color from top to bottom) and extrapolated $N=2,4$ (black, respectively solid and dashed) and  T1. We also show the BAM $r=100$ case in dashed blue. }
\label{fig:epsilonsSXS}
\end{figure*}

We show the values of the $\epsilon_\lm$ as a function of frequency in Fig.~\ref{fig:epsilonsSXS}. As expected from our analysis of the 
$\Lambda_{\ell m}$, the errors are again dominated by the finite wave extraction radius. The largest radii still differ significantly from the extrapolated waveforms, for which the agreement between NR and PN is of the order of only $1^{\circ}$. Again, the behavior of the  $(2,1)$ mode is typical for the $(\ell,\ell-1)$ modes, while the $(\ell,\ell)$ modes behave similar to the $(3,3)$ shown here. We note that the finite radius errors are strongly dependent on the gauge conditions used in the numerical relativity code.%\textcolor{red}{It would be good to have N=4 and 2 in both plots. and the first plot should reach a bit lower, say to -5 degrees so that the extrapolated curve is away for the axis and clearly visible. Would it be possible to add BAM r=100 for the frequencies where it is available? Maybe in this case one should also put SXS r=100}

\begin{figure*}[htbp]
\includegraphics[width=0.85\columnwidth]{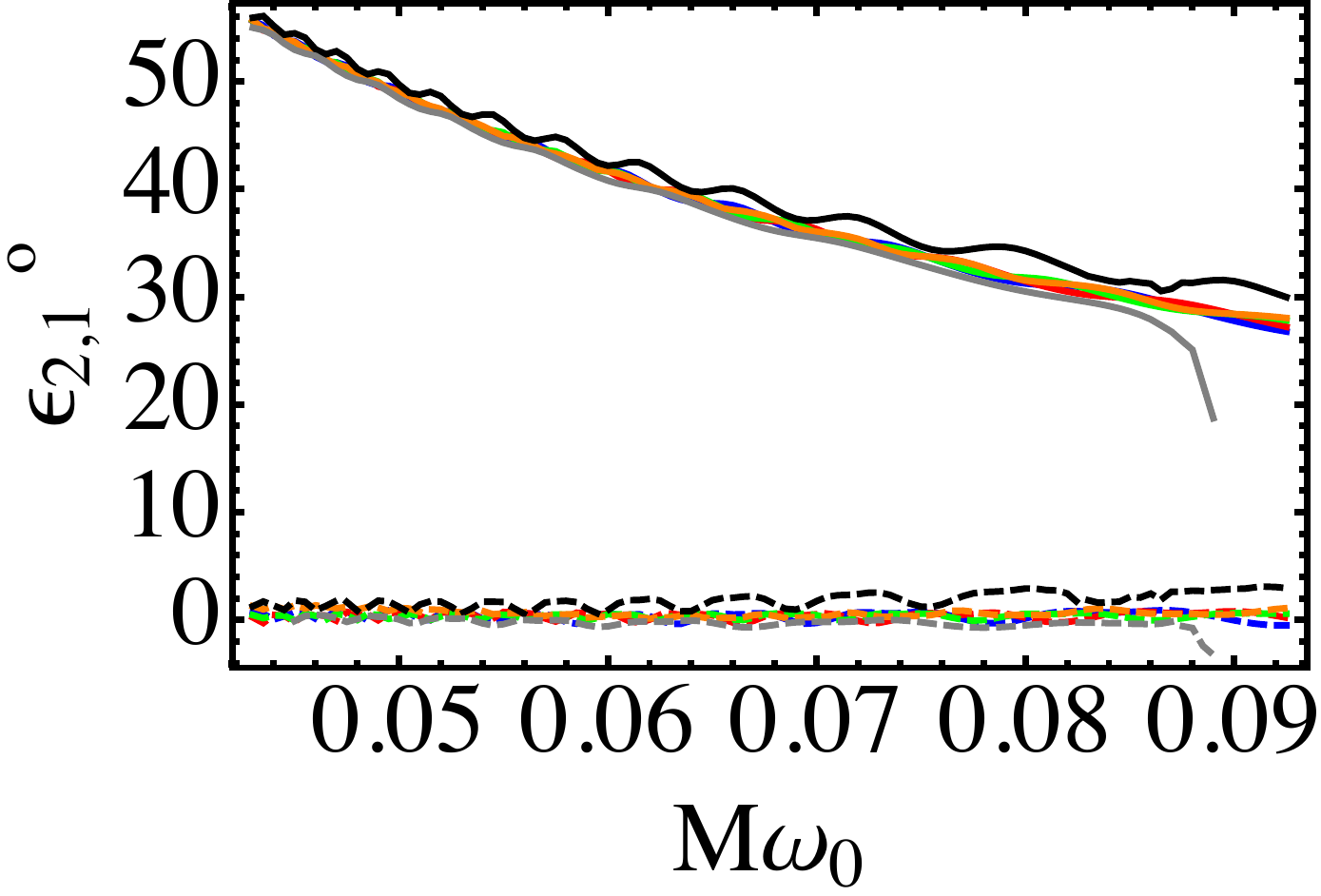}
\includegraphics[width=1.15\columnwidth]{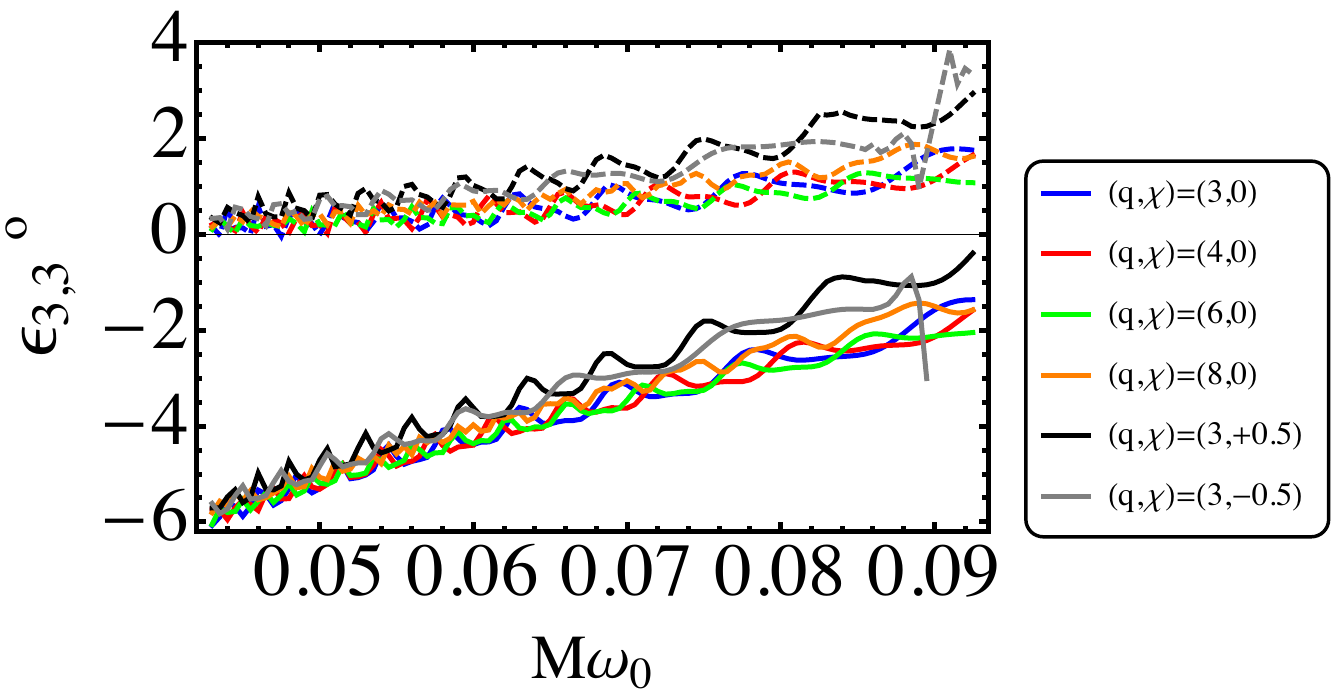}
\caption{$\epsilon_\lm$ values for several SXS catalog systems for  $r=100M$ and extrapolated $N=4$ data (solid and dashed respectively). Non-spinning systems are shown in color while we use black and grey for positive and negative spins respectively. Note that the values of $\epsilon_\lm$ are almost equal for all the systems, specially among the non spinning ones.
%We cut the $(q,\chi)=(3,-0.5)$ curves at $M\omega_0=0.090$ due to the PN phase not being accurate enough anymore, affecting this the estimation of $\varphi_0$.
}
\label{fig:epsilonsSXSq}
\end{figure*}

We have also analysed the non-spinning $q=(3,4,6)$ and the aligned-spin $(q=3,\chi=\pm 0.5)$ cases. We find  that for systems with different mass ratios and spins the quantities $(r_\lm, \Lambda_\lm, \epsilon_\lm)$ behave qualitatively the same as function of extraction radius. Furthermore, in Fig.\ref{fig:epsilonsSXSq} we see that the values of $\epsilon_\lm$ for all the mentioned cases are almost the same for equal extraction radius along all the studied frequency range. This suggests that the influence on the $\epsilon_\lm$ of the extraction radius $r$ of NR waveforms depends on the specific parameters of the simulated systems rather weakly.

\section{Effect of NR extraction radius and extrapolation on the match}
\label{sec:MatchResults}

In the previous sections, we have investigated the disagreement between PN and NR in the region where we align and attach them. The discrepancies we observed illustrate inaccuracies in both waveforms at the typical frequencies where we perform the hybridization which will contribute to the global error budget of the hybrid. However, the accuracy of the full hybrid is of course also affected by the details of how these discrepancies are smoothed in the hybridization window and most importantly by the intrinsic error of the PN and the NR portions before and after the hybridization region. There, the $r_{\ell m}$ and $\epsilon_{\ell m}$ do not inform us about the accuracy. For instance, it could be that the spurious relative dephasings between the NR modes that we have observed may have disappeared around the merger where the higher modes are most important for matches.
It is therefore not possible to directly translate the values we observed for the residual phase and amplitude disagreements into an overall error of the final hybrid.

In this section, we take a first step towards quantifying the error budget of the full hybrid in terms of quantities useful for data analysis applications. As usual, we will replace the unanswerable question of how much the hybrid differs from the true signal by a study of the typical mismatch that one obtains when one varies the different ingredients in the procedure. A systematic study, which properly addresses the different requirements for the detection and parameter estimation problems across a significant portion of the black hole parameter space,  is beyond the scope of this paper.
Instead, we restrict the scope of this study to understanding the effect of the extraction radius of the NR waves (and their extrapolation) on the match. In doing this, we do not claim that this is the main source of inaccuracy for the hybrid: using for instance a PN approximant that predicts a phase evolution very different from the NR one (see e.g.  T4 in Fig.~\ref{fig:psi0q18NS}) will certainly lead to larger mismatches. Rather, motivated by the observations of the previous section that suggest that extraction radius can have a strong effect on the agreement between PN and NR in the late inspiral, we illustrate here how much of an effect it makes on the match between full hybrids. Comparing the results to other studies in the literature can give a first impression of the relative contribution of different imprecisions for data analysis applications and guide more systematic future investigations.

We illustrate our results with the $q=8$ non-spinning case from the SXS catalog and use the 2015 early Advanced LIGO noise curve \cite{Aasi:2014tra}. We hybridize with  T1 (the PN approximant that gives the smallest secular trend between the PN and the NR $(2,2)$ mode frequencies for this case) hybridized at $\omega_0=0.073$. With this choice, and given the length of our blending windows, the NR $(2,2)$ mode covers the entire instrumental band (say starting from $20$ Hz) for total masses larger than $\sim120 M_{\odot}$. Here we have chosen to hybridize at a relatively large frequency to facilitate comparisons with shorter NR simulations.
 We focus for now on the highest resolution available (namely \verb+Lev. 5+) and on the waves extracted at $r=100$M (innermost radius available) and $307$M (outermost one) as well as those extrapolated to null infinity using $N=2$ and $N=4$ polynomial orders. We denote these waveforms $h_{100}$, $h_{307}$, $h_{N=2}$ and $h_{N=4}$ respectively.

\begin{figure*}[htbp]
\centering
\includegraphics[width=1.1\columnwidth]{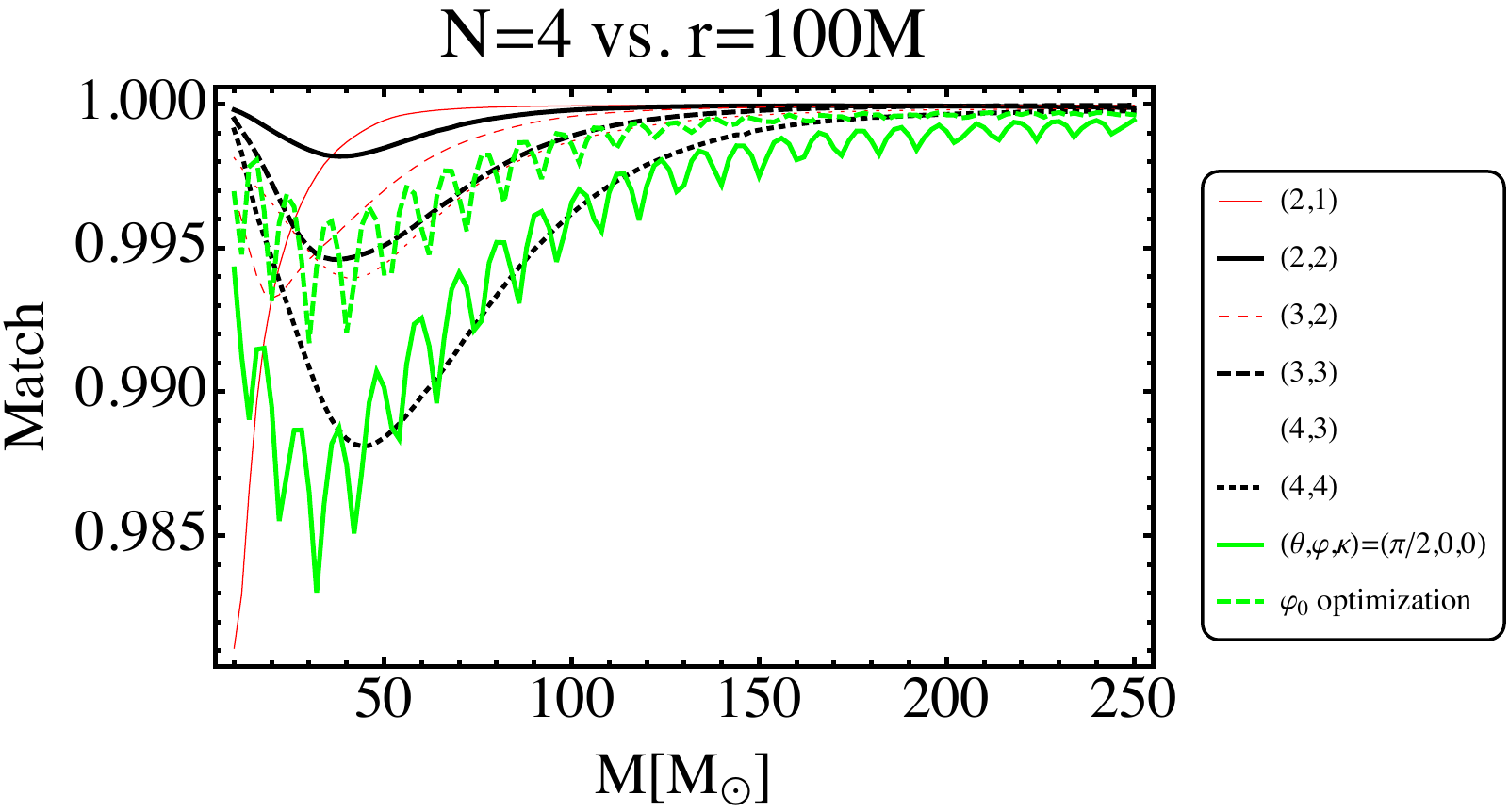}
\includegraphics[width=.9\columnwidth]{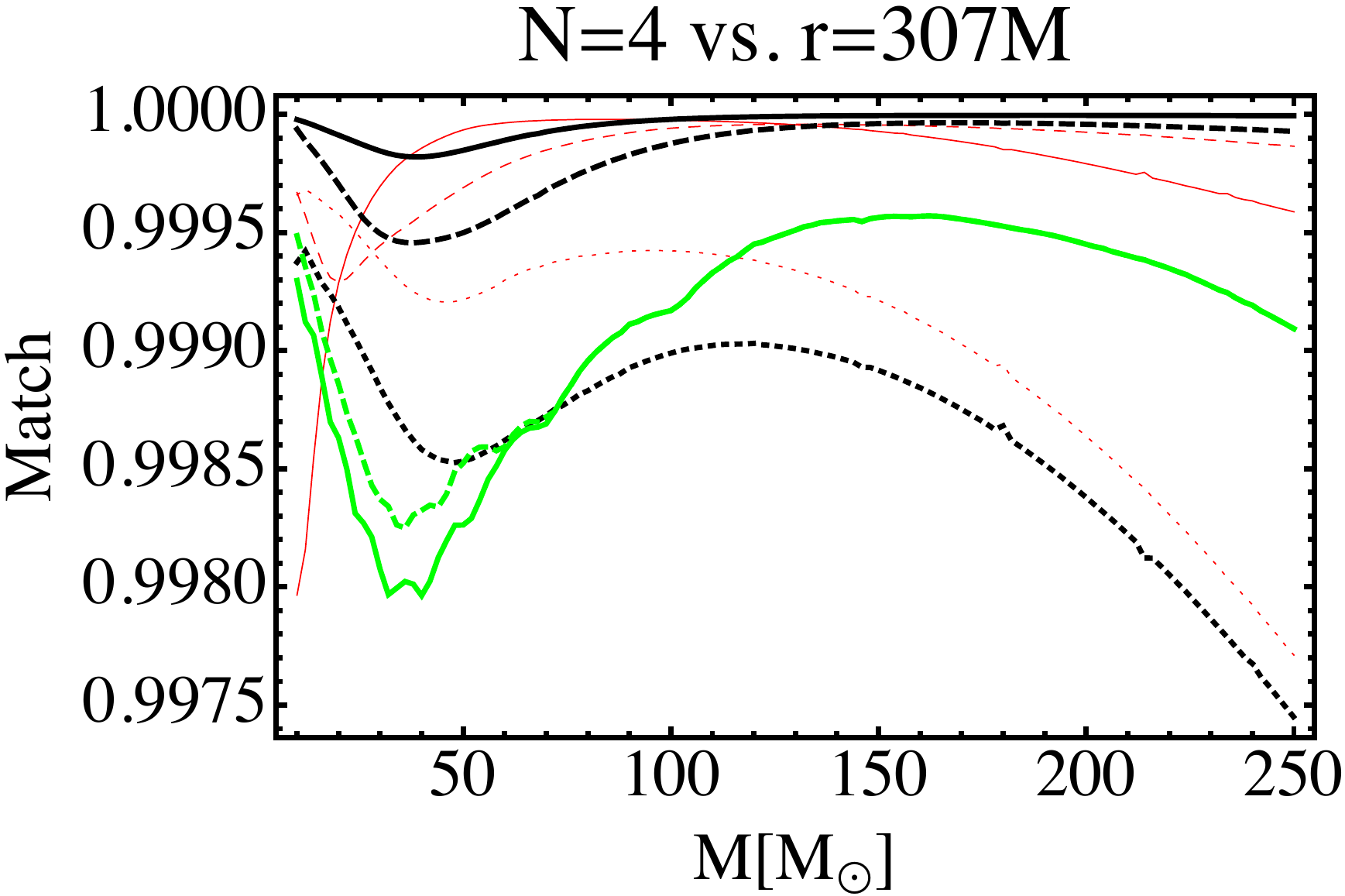}
\includegraphics[width=.9\columnwidth]{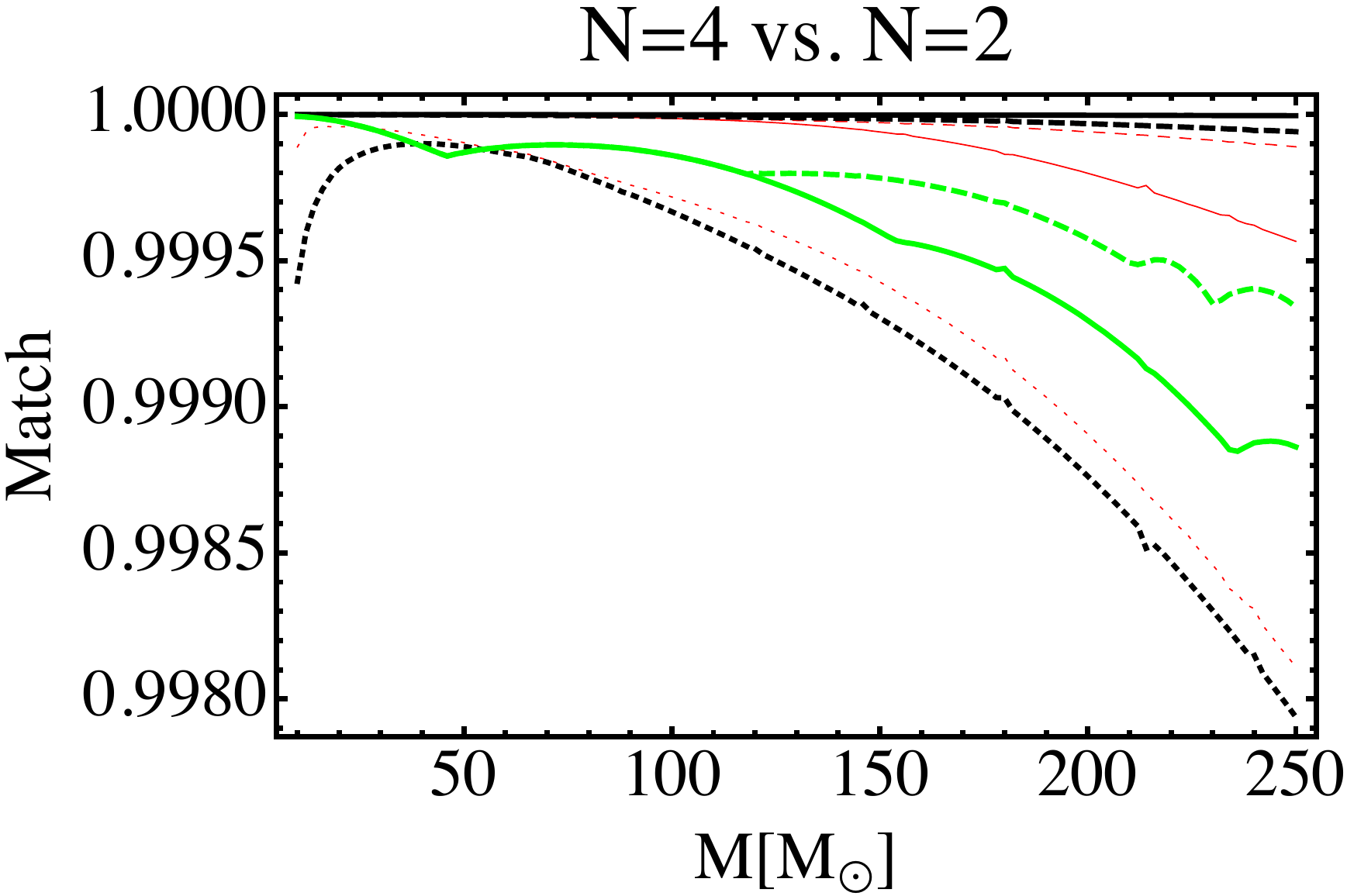}
\hspace{.2\columnwidth}
\includegraphics[width=.9\columnwidth]{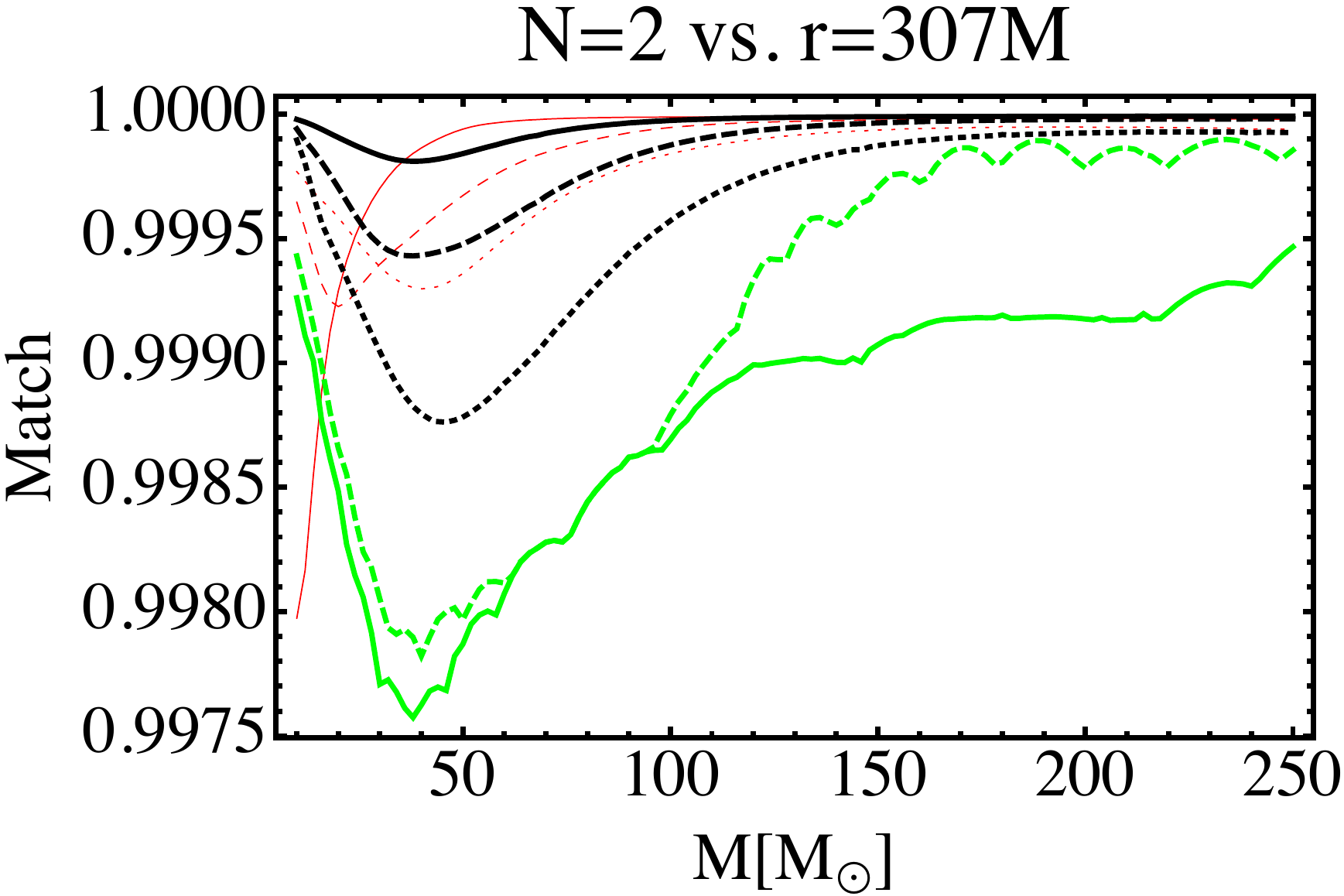}
\caption{Match of individual modes (optimized over phase and time) and full waveforms at $(\theta,\varphi,\psi)=(\pi/2,0,0)$  optimized only over time (in solid green) and time and phase (dashed green) for $q=8$ non-spinning hybrid waveforms built out of T1 and SXS NR waveforms produced during a single simulation but either extracted at finite radius $r=100M$ and $r=307M$ or extrapolated with polynomial orders $N=2$ and $N=4$.}
\label{fig:matchSXS}
\end{figure*}

Our hybridization procedure applied to each of these numerical waveforms yields a hybrid $h_X(\theta,\varphi,\kappa)$ (with the notation of Eq.~\eqref{eq:hthetaphikappa}) whose modes are defined in Eq.~\eqref{eq:hom} and slightly different values for the shift $\varphi_0$ needed to adjust conventions \footnote{We find $\varphi_0^X-\varphi_0^{N=4}= .04^\circ,\, 7.19^\circ,\, 21.87^\circ$ for $X=(N=2),\,307,\,100$ respectively. We find the same $\psi_0$ for all these cases.} between PN and NR. This is of course a mere consequence of the fact that our procedure to infer the differences in conventions is affected by the inaccuracies in the original waveforms (here we know that the conventions are the same for all the NR waveforms that we consider).
%\textcolor{red}{(In fact, Juan, not to put in the text but just for us, could you add here the values for each one? I want to have an idea of the spread. Thanks.)}\juan{$N=2:56.54;N=4:56.50;r=307:63.69;r=100:78,37$ degrees. The reason for hybridizing so late is that we aim to see the effect of the hybridization region in cases where higher modes are important: i.e, we wanted the hybridiation region to have effects at M=70,78 and also that not for all codes it is possible to hybridize at $M\omega \leq 0.043 $}
Because in the presence of higher modes, the dependence on $\varphi$ is non trivial, one has to be careful to compare waveforms at the same physical sky orientation (or optimize over it depending on the application). In the general case, if two hybrids are built out of different PN pieces and different NR pieces, comparing at the \emph{same} angle $\varphi$ makes no sense in principle since the definitions of the origin of the azimuthal angle are a priori independent. In the present case however, $\varphi$ has the same meaning for all the numerical waveforms that we consider (and therefore, with our choice of applying the $\varphi_0$ rotation to the PN part in Eq.~\eqref{eq:hom}, for the hybrids) and it makes sense to compute
\be
\label{eq:matchsamephi}
\max_{t_0} {\cal O}\left[h_1(\pi/2,0,0),h_2(\pi/2,0,0)\right],
\ee
i.e. the overlap optimized over time-shifts only of the two waveforms at the same (source) sky location $(\theta=\pi/2,\varphi=0)$ chosen so that the higher modes contribute significantly to the full waveform and at the arbitrary effective polarization $\kappa=0$. This is the quantity plotted in plain green in Fig.~\ref{fig:matchSXS}. Note however that the early inspiral tails of our hybrids now differ by a shift in $\varphi$, i.e. despite the fact that we are using the same PN input in all our hybrids, the quantity defined in Eq.~\eqref{eq:matchsamephi} will not go exactly to $1$ in the limit of small masses where only the PN tail is in band\footnote{An equally valid point of view would be to modify Eq.~\eqref{eq:hom} by applying the $\varphi_0$ shift to the NR portion, i.e. redefining the conventions of the NR waveforms to reproduce those of the PN (as always up to errors in the determination of this convention shift). Then, when computed using the waveforms evaluated at the same $\varphi$, the match goes to 1 for small enough masses since we are comparing the original waveform against itself but differs from the match between the pure NR waveforms at masses large enough that the hybridization region is out of band.}. Often in data-analysis applications, it is also interesting to optimize this match over $\varphi$, in which case all the subtleties of the exact definition of the azimuthal angle of the hybrid become irrelevant. We therefore also plot in dashed-green the quantity
\be
\label{eq:matchoptimizephi}
\max_{t_0,\varphi} {\cal O}\left[h_1(\pi/2,0,0),h_2(\pi/2,\varphi,0)\right]
\ee
in the figures below.

Fig.~\ref{fig:matchSXS} (beware of the different scale for the top left panel) displays the result of this study for various couples of waveforms. In each panel, we additionally show the overlap between the individual modes (in black and red) optimized over time and phase to check to what extent the modes of both hybrids agree if we allow ourselves to align them one by one independently. We first focus on the top panel. In order to interpret these plots, one should keep in mind that in the large mass limit, the hybridization window is pushed to lower frequencies than those accessible to the instrument and this becomes a pure NR comparison. In this region, the typical mismatch between $h_{N=4}$ and $h_{100}$ is of a few $0.1\%$ and goes below $0.1\%$ for $h_{307}$. As we move to smaller masses and the hybridization region enters the instrumental band, the match degrades and reaches a minimum around $40M_{\odot}$: while the mismatch there is above $1\%$ in the $R=100$ vs $N=4$ case (this gets reduced by a factor $\sim2$ after $\varphi$-optimization), it remains as low as $0.2\%$ when using $R=307$. At even lower masses, the comparison becomes dominated by the PN tail which is identical for both waveforms up to the $\varphi$ shift discussed above and the match grows again (the optimized one going to 1 exactly in the low mass limit). 

From this comparison where the $N=4$ waveforms has been used as a reference, one can argue that for the NR data set studied here, extracting the waves in the SpEC code at radii of $\mathcal{O}(300\, M)$ (as typically available in the SXS catalog) is sufficient to control the error to the $\sim 0.1\%$ level in terms of mismatch. 

To check the effect of extrapolation orders on this study, in the lower right panel we reproduce the upper right one but use $N=2$ instead of $N=4$. While the general behavior and scale remains essentially unchanged, we note that the $(4,4)$ and $(4,3)$ modes which were significantly disagreeing at high frequencies and dominating the total mismatch between $N=4$ and $R=307$ now give much higher matches at high masses (note also that after optimization over $\varphi$ the matches between the hybrid become very close to $1$).  This is consistent with the Ref.~\cite{Boyle:2009vi}, where it is also observed that different orders may show varying performance during the evolution, and higher orders may in particular be problematic during merger-ringdown.
Except for these discrepancies at high mass which come from the presence of unphysical features in these two $N=4$ modes and which remain on the order of $\sim 0.1\%$, the order of extrapolation therefore does not seem to affect our previous conclusion, as also illustrated by the lower left panel in which we directly compare the hybrids built with $N=4$ and $N=2$.

\begin{figure}[htbp]
\centering
\includegraphics[width=\columnwidth]{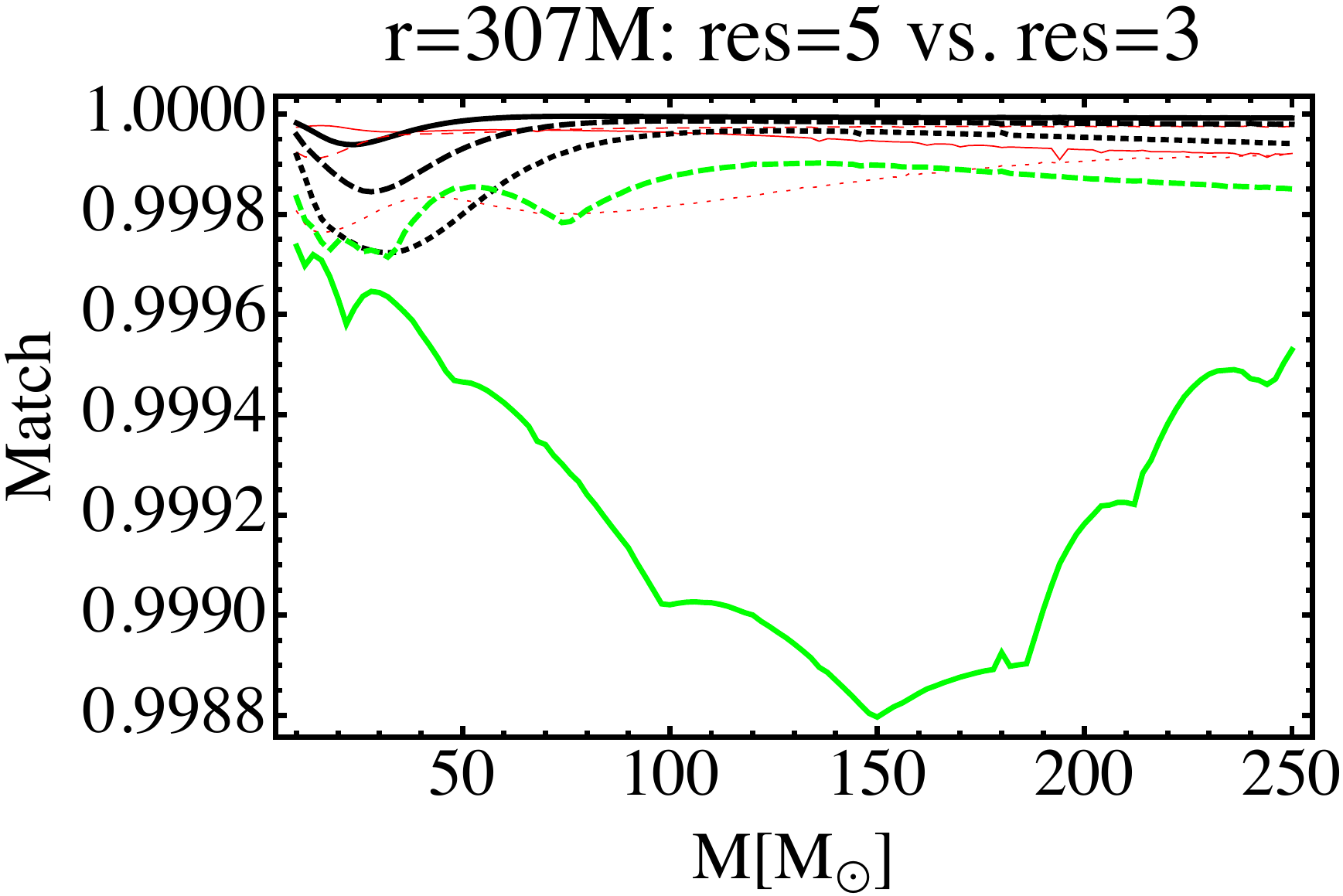}
\caption{Match as in Fig.~\ref{fig:matchSXS}, but between hybrids constructed from numerical waveforms produced at different numerical resolutions instead of extraction radii. We use the highest available resolution level (5) and a lower resolution level (3) for $R=307M$.}
\label{fig:matchSXS_ref}
\end{figure}

Finally, we perform a similar study to compare the effect of numerical resolution on the match to the effect from different extraction radii. In Fig.~\ref{fig:matchSXS_ref} we plot matches as in Fig.~\ref{fig:matchSXS}, but using hybrids constructed from numerical waveforms produced at different numerical resolutions instead of extraction radii. For the faster than polynomial convergence exhibited by spectral codes such as SpEC  one typically quotes the difference between the highest and next-highest resolution as an error estimate, see e.g.~\cite{Hinder:2013oqa} for a comparative discussion of the error analysis performed in spectral and finite-difference numerical relativity codes. Here we conservatively use  the highest available resolution level (5) and a resolution two levels coarser (3), and we find mismatches smaller or at comparable level that the mismatches shown in Fig.~\ref{fig:matchSXS} resulting from  finite extraction radius. In both cases, the mismatches are certainly not larger than what can be expected in waveform models such as \cite{Ajith:2007qp,Santamaria:2010yb,Buonanno:1998gg,Taracchini:2013rva}. In addition, any actual signal searches or parameter estimation calculations will be based on discrete or continuous waveform families and matches will effectively be optimized over all physical parameters.

\section{Discussion}

Gravitational wave searches based on the data from ground based detectors have to date been based on template banks which only contain the dominant $(\ell,|m|)=(2,2)$ spherical harmonic modes of the signal. Several authors have recently started to evaluate the impact of this restriction on detection and parameter estimation \cite{Pekowsky:2012sr,Brown:2012nn,Capano:2013raa,Varma:2014jxa}. Such studies have used multi-mode effective-one-body models and hybrid waveforms directly constructed from the numerical data and an inspiral model. In addition, such multi-mode hybrids will facilitate the extension of phenomenological  \cite{Santamaria:2010yb,Hannam:2013oca}  and similar waveform models to the multi-mode case.

The goal of the present paper has been to achieve a better understanding of the construction and properties of such hybrids for complete waveforms. To this end, we have first compared the subdominant modes of non-precessing (and mostly non-spinning) NR data sets with PN expressions.
We have first described the ambiguities in such a comparison, and described how to determine the three parameters that parameterise different conventions in the literature and available data sets: a time shift 
$\tau$, rotation around the orbital angular momentum by an angle $\varphi_0$, and a polarisation angle $\psi_0$.
For non-precessing systems it can be assumed that  $\psi_0 \in \{0,\pi\}$. We have given a prescription to determine  $\psi_0$ from numerical data, although in principle it should be deducible from the description of a NR code or PN calculation.

The presence of PN truncation errors and of numerical errors in NR waveform descriptions results in amplitude and phase discrepancies during the late inspiral (where we can compare them), which cannot be compensated by the choice of $\tau, \varphi_0$ and possibly $\psi_0$, and thus some choices have to be made to determine the values of these quantities for a given multi-mode waveform.  In our construction we first determine the time shift $\tau$ from the (2,2) mode alone, and then determine the value of $\varphi_0$ from two modes, one of which is typically (2,2).
We have then studied the deviations between the PN result and NR data sets as a function of matching time.

Regarding NR amplitude errors, we find a strong dependence on the numerical code used, which we attribute to the difference in coordinate gauge conditions. With the exception of the $(3,2)$-mode, e.g. $r=100$ BAM data are significantly closer to the extrapolated result than the corresponding SXS $r=100$ curve.
We also note that, for any given finite extraction radius, the error becomes larger at lower frequencies. This is the expected consequence of the fact that as the frequency increases (or equivalently as their wavelength decreases), the wave zone (defined by $r\gg\lambda$) extends to smaller radii.
We find good agreement between post-Newtonian amplitude and those extrapolated to infinity from several
extraction radii in a numerical relativity calculation for the $(2,1)$, $(2,2)$ and $(3,2)$ modes, but significantly larger errors of up to a few $10\%$ for the modes $(3,3)$, $(4,3)$, $(4,4)$ and $(5,5)$, and we find that these deviations are consistent with the spread between PN results at different orders.

Regarding the phase differences  $\epsilon_\lm$,  we find excellent agreement on the order of $1^{\circ}$ of PN results with numerical relativity waveforms extrapolated to null infinity, while deviations at finite radius can be as large as tens of degrees for $(\ell,\ell-1)$ modes. On the contrary, for the $(\ell,\ell)$ modes, the larger finite radii curves only differ from the extrapolated one by of the order of $2$ degrees. In particular, we find that the influence on the $\epsilon_\lm$ values of the extraction radius $r$ of NR waveforms depends on the specific parameters of the simulated systems only rather weakly. Our results imply that during the inspiral the complex part of the standard PN waveform amplitudes can be taken as the correct value for practical purposes. This yields in particular a convenient 
test for finite radius errors in numerical relativity (since a practically exact result is known), and may serve to determine favorable coordinate gauge conditions for wave extraction.

A systematic study of the effect of the phase and amplitude errors which we discuss on gravitational wave data analysis, both regarding  the detection problem and parameter estimation is beyond the scope of this paper.
Instead we have performed simple match calculations with the initial and design sensitivity advanced LIGO noise curve, evaluating the match  between waveforms resulting from different extraction radii or numerical resolutions. This simplistic study can serve for comparisons due to mismatches resulting from other effects, such as the choice of PN approximant, or waveform modelling errors, and thus give a first impression of the relative importance of different types of waveform imprecisions, and help guide more detailed investigations. This part of our study overlaps with other investigations such as 
\cite{Boyle:2009vi,Pollney:2009yz,Taylor:2013zia} regarding the errors in finite radius and extrapolated waveforms. Our results appear consistent with previous work, but add the aspect of considering full hybrids and investigating how the match varies when the hybridization region is in band and puts the focus on errors in higher modes.

In Sec.~\ref{sec:errorsources} we have seen in particular that the extraction radius can lead to significant phase and amplitude errors. Corresponding mismatches as discussed in Sec.~\ref{sec:MatchResults} are however roughly at the level of $0.1\%$ for the cases we have considered.  Mismatches at this level appear at least negligible for GW searches. We conclude with the remark that a systematic understanding of multi-mode waveform errors for parameter estimation necessarily needs to take into account precession effects.

\section{Acknowlegdements}
We thank P. Ajith for discussions.
JCB, SH, AMS and in part AB were supported the Spanish Ministry of Economy and Competitiveness grants
FPA2010-16495, CSD2009-00064, FPA2013-41042-P, European Union FEDER funds and Conselleria d'Economia i Competitivitat del Govern de les Illes Balears.
MH was supported by STFC grants ST/H008438/1 and ST/I001085/1.
MP was supported by STFC grant ST/I001085/1. This document has been assigned LIGO Laboratory document number LIGO-P1500004.

\bibliography{HWbib}

\end{document}